\documentclass[10pt, journal]{IEEEtran}
\IEEEoverridecommandlockouts

\usepackage[pdftex]{graphicx}

\usepackage{epstopdf}

\usepackage{amsthm}
\usepackage{amssymb}
\usepackage{mathptmx}
\usepackage{array}
\usepackage[cmex10]{amsmath}
\usepackage{amsfonts}
\usepackage{comment}
\usepackage{xcolor}
\usepackage[boxed,ruled,vlined,linesnumbered,commentsnumbered, noresetcount]{algorithm2e}
\usepackage{algorithmic}
\usepackage{setspace}
\usepackage{mathtools}
\usepackage{cite}
\usepackage{flushend}
\usepackage{color}
\usepackage{threeparttable}
\usepackage{diagbox}
\usepackage{adjustbox}
\usepackage{booktabs}
\usepackage{multirow, makecell}
\setcellgapes{3pt}

\usepackage[utf8x]{inputenc}



\usepackage{subcaption}

\usepackage{pifont}

\newcolumntype{P}[1]{>{\centering\arraybackslash}p{#1}}
\newcolumntype{M}[1]{>{\centering\arraybackslash}m{#1}}

\begin{document}

\title{Wireless Communications for Smart Manufacturing and Industrial IoT: Existing Technologies, 5G, and Beyond}

\author{Md. Noor-A-Rahim,  Jobish John, Fadhil Firyaguna, Dimitrios Zorbas, Hafiz Husnain Raza Sherazi,  Sergii Kushch, Eoin O’Connell,  Dirk Pesch, Brendan O’Flynn, Martin Hayes,  and Eddie Armstrong

\thanks{Md. Noor-A-Rahim,  Jobish John, Fadhil Firyaguna,  and Dirk Pesch are with the  School of Computer Science \& IT, University College Cork,  Ireland.  (E-mail: {\tt \{m.rahim, j.john,f.firyaguna, d.pesch\}@cs.ucc.ie}). 

Dimitrios Zorbas is with the School of Engineering \& Digital Sciences, Nazarbayev University, Nur-Sultan, Kazakhstan. (E-mail: {\tt dimzorbas@ieee.org}).

Hafiz Husnain Raza Shaerazi is with the School of Computing \& Engineering, University of West London, London,  United Kingdom. (E-mail: {\tt sherazi@uwl.ac.uk}).

Sergii Kushch, Eoin O’Connell, and Martin Hayes are  with the Department of Electronic \& Computer Engineering, University of Limerick, Limerick, Ireland. (E-mail: {\tt \{sergii.kushch, eoin.oconnell, martin.j.hayes\} @ul.ie}). 

Brendan O’Flynn is with the Tyndall National Institute, University College Cork, Cork, Ireland. (E-mail: {\tt brendan.oflynn@tyndall.ie}).

Eddie Armstrong is with the Johnson \& Johnson, Ireland. (E-mail: {\tt earmstr1@its.jnj.com}).


}
}

\maketitle

\begin{abstract}
Smart manufacturing is a vision and major driver for change in industrial environments. The goal of smart manufacturing is to optimize manufacturing processes through constantly monitoring and adapting processes towards more efficient and personalised manufacturing. This requires and relies on technologies for connected machines incorporating a variety of computation, sensing, actuation, and machine to machine communications modalities. As such, understanding the change towards smart manufacturing requires knowledge of the enabling technologies, their applications in real world scenarios and the communications protocols that they rely on. This paper presents an extensive review of wireless machine to machine communication protocols currently applied in manufacturing environments and provides a comprehensive review of the associated use cases whilst defining their expected impact on the future of smart manufacturing. Based on the review, we point out a number of open challenges and directions for future research.


\end{abstract}

\IEEEpeerreviewmaketitle





\section{Introduction}

Increased connectivity and collaboration between workers, equipment, processes and products through industrial IoT systems and machine to machine communication enables an important aspect of Industry 4.0: the capacity to increase the creation of actionable knowledge from the data being produced by both the production and management systems within a manufacturing site through the introduction of integrated smart digital technologies. An example of the consequence of this pursuit of digital transformation is in the aim of manufacturers to deliver personalized mass customized products at an equivalent price point to mass-produced goods. To achieve this goal, it is necessary to bring more automation and productivity to factory operations wherever possible in a low volume setting. This objective requires an optimal information flow between supply chain, engineering, sales and operations, all of which require reliable, flexible network connectivity. This challenge has been evident in the drive towards the digitalization agenda and the need to increase the flexibility of automation systems and processes towards the aims of Industry 4.0 \cite{7019732}.

As part of this trend, many companies are evaluating the transition from wired to wireless machine-to-machine (M2M) communications and the embedded Internet of Things (IoT) to enable the rapid and inexpensive addition of  smart sensor technologies to legacy machinery, enhancing the ability to quickly reconfigure manufacturing lines for ``batch size of one" operation, to enable mobile robot operation, integrate communication across both factory and supply chain operations, and to enhance real-time on-product tracking and decision-making on the factory floor. This has been leading to a growth in the deployment of IoT systems across many sectors, with the expected number of devices reaching 29.4 billion by 2030 \cite{statista}. Deploying wireless technologies in a factory or a manufacturing environment is a complex undertaking due to a number of factors such as the very nature of the manufacturing environment itself, whether it is indoor or outdoor, and the corresponding challenges for radio propagation characteristics in these diverse environments. Engineers also need to be cognisant of the susceptibility of wireless communications to noise and electromagnetic interference, as well as the power requirements for wireless devices and the general lack of deployment experience of wireless technologies in manufacturing environments. Networking large numbers of sensors, actuators, effectors and machine control systems in a dynamic factory environment can be achieved in many ways, such as by using a wide variety of standards and non-standards based wireless networking technologies. 

The specific requirements of industrial IoT and automation systems for highly reliable, low latency connectivity has meant that wireless protocols have had to become increasingly more technologically advanced whilst remaining easy to use. This aim has led to a rapidly growing catalogue of wireless protocols and IoT standards, technologies and platforms targeting this new ecosystem, which engineers and technologists need to understand and integrate into their manufacturing processes. Choosing the optimal standards and IoT platforms and systems is a constant challenge for system designers trying to keep pace with the rate of development. They must be cognisant of the potential interoperability issues between various IoT options, associated M2M communications standards and capabilities, and support a lifetime of specific technology choices, whether they are open or proprietary solutions. As such, there is a requirement to compile information on the range and capabilities of wireless technologies for industrial environments to compare and contrast the wide range of wireless protocols currently available. In this paper, we present a comprehensive survey on the current and future wireless technologies for industrial IoT and discuss the applicability and limitations for a range of typical smart manufacturing use cases.

\begin{figure*}[h]
    \centering
    \includegraphics[width = 12cm]{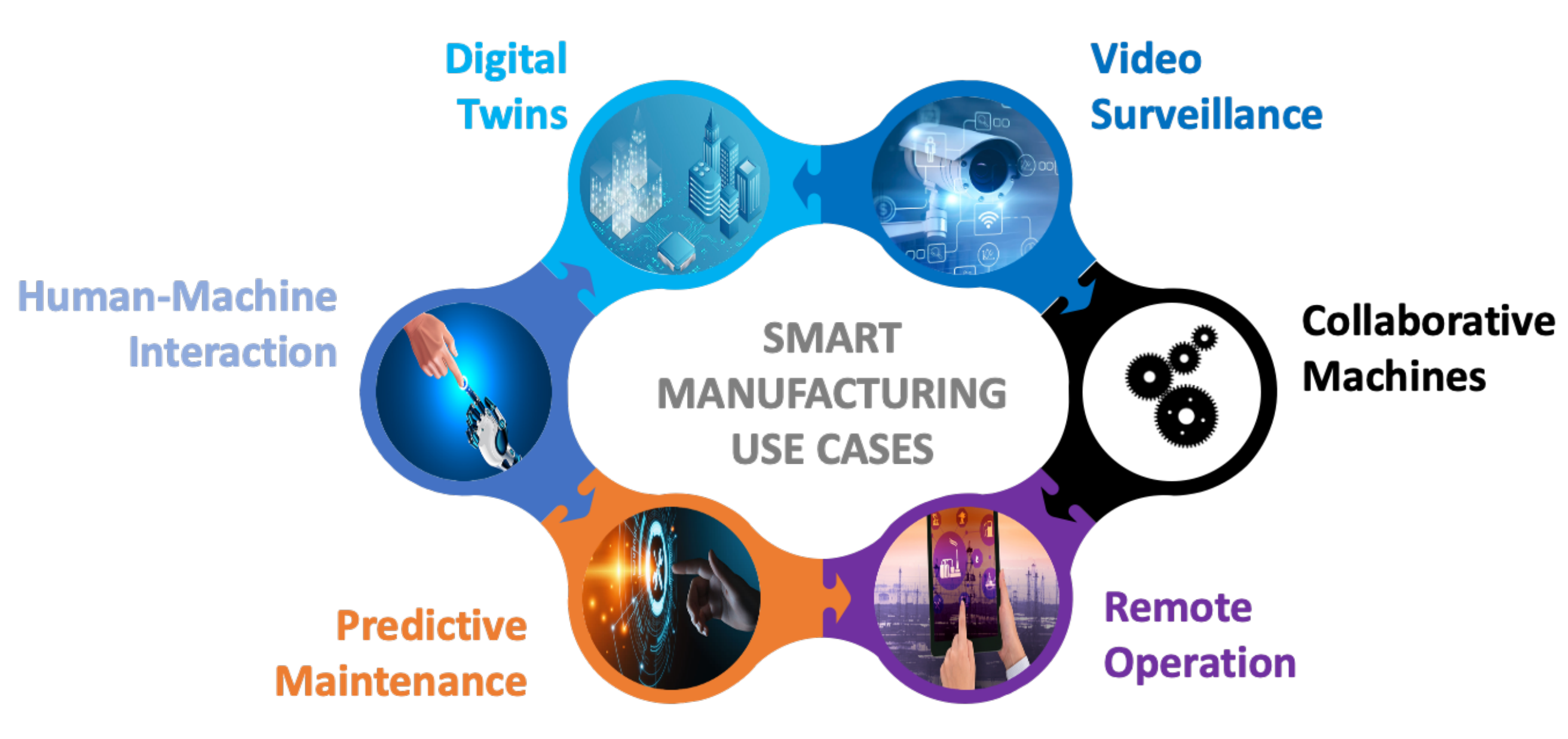}
    \caption{Examples of smart manufacturing use cases.}
    \label{fig:usecases}
\end{figure*}

\subsection{Related Survey Papers}


Over the  past  decade, a  number of survey papers  have  been  published aimed  at  addressing   various   aspects   of  wireless technologies in industrial automation. For example,  an overview of industrial wireless networks and their challenges is surveyed in \cite{Willig2005,Wollschlaeger2017,Park2018,DePellegrini2006,Li2017}. A comparative study  on architecture and protocol design, which considered  four prominent  industrial wireless technologies based on the  IEEE 802.15.4 standard (e.g. ZigBee, WirelessHART, ISA100.11a and WIA-PA),  was presented in \cite{Wang_2016}.  Similar to \cite{Wang_2016}, the protocol suitability for medium access control (MAC), routing and transport functions  was explored in \cite{Kumar_2014}. A comparison of field bus technologies, industrial Ethernet and wireless solutions was carried out in \cite{Frenzel2013}, and the connection of monitoring and control operations for Wi-Fi, Bluetooth, ZigBee and WirelessHART technologies is also considered. Requirements for industrial communication networks for process automation at the level of field devices were considered in \cite{Ikram2016}. Authors in \cite{Sanchez-Iborra2016} have focused on the state-of-the-art in Low-Power Wide-Area Network (LPWAN) technologies for industrial environments.  Authors in \cite{Christin2010} surveyed the IEEE 802.15.1 and IEEE 802.15.4-based technologies in an industrial space from the perspective of security and quality of service. More recently, \cite{Seferagic2020} has provided a comparative performance evaluation of  traditional industrial wireless technologies in terms of different performance metrics including latency, reliability, scalability, energy consumption, data rate and coverage.  While the existing surveys cover the above-mentioned traditional industrial wireless technologies (for instance of IEEE 802.15.4),  the majority of them do not comprehensively cover the entire industrial wireless problem space. Moreover, they do not discuss the most recent technologies being developed in the context of smart manufacturing. To the best of our knowledge, none of the existing works discuss the potentials for emerging (e.g. 5G) and future  (e.g., 6G) technologies in the industrial sector. With the rapid research and development being carried out in relevant emerging technologies, an up-to-date survey is required to establish and analyze the applicability of emerging and future communications technologies in the context of smart manufacturing. 

\subsection{Contributions}
Motivated by the importance of wireless communications in Industry 4.0 applications articulated in the previous section, and to stimulate further research and innovation in the area of wireless smart manufacturing, this paper aims to bridge the existing knowledge gap in the area by presenting an overview of the different wireless protocols that are currently, or soon to be, available for industrial IoT applications. The paper will also explore the specific use cases that are capitalizing on M2M communications and evaluate the technologies that would need to be deployed to facilitate the successful implementation of the aforementioned use cases. The contributions of this paper are summarized as follows:

\begin{itemize}
    \item A study of use cases in smart manufacturing and their wireless communications requirements.
    \item A detailed survey of existing wireless technologies in smart manufacturing and their applicability in different industrial applications.
    \item A study of emerging and future wireless technologies and their potential  applications in industrial environments.
    \item The provision of future directions and challenges for wireless smart factories in areas such as wireless energy harvesting and mm-Wave communications.
\end{itemize}

\subsection{Paper Organization}
The remainder of this paper is organized as follows. In Section~\ref{sec:use_cases}, we discuss some high level manufacturing use cases  and their requirements in terms of performance of M2M communications. The specifications of existing wireless  networking  technologies targeting industrial applications are presented in Section~\ref{sec:exis_tech} along with their applicability in a  wide  range  of  smart manufacturing use cases. In Section~\ref{sec:Emerging_Tech}, we present  emerging and future  wireless technologies and their potential applications in the industrial space. In Section~\ref{sec:Future_Works}, a wide range of  important  future  directions for wireless technologies in smart manufacturing are discussed. Finally, in Section~\ref{sec:conclusion}, we summarize the current state-of-the-art and draw conclusions.

\section{Smart Manufacturing Use Cases and Requirements}
\label{sec:use_cases}
Over the past couple of decades, industry (in a variety of sectors including automotive, aerospace, equipment manufacturing, or oil \& gas) are integrating Internet of Things (IoT) technologies into their manufacturing infrastructure.  This is due to IoT's capabilities to help companies optimize production processes, reduce delivery times and enable substantial reduction in operational costs, while improving production efficiency. There are numerous potential use cases for industrial IoT ranging from asset tracking and management to logistics, security, and customer servicing. In this section, some high level manufacturing use cases (as illustrated in Fig.~\ref{fig:usecases}) are identified and their requirements in terms of performance for M2M communications and IoT are outlined. 

\subsection{Smart Manufacturing Use-cases}
\subsubsection{Cyber-Physical Production Systems}
Being a key enabler in the evolution towards Industry 4.0, the concept of Cyber-Physical Production Systems (CPPS) \cite{uhlemann2017digital} is well established and is characterized by the integration of computational processes with physical processes implemented throughout a smart production environment. This implies that physical processes are consistently examined and controlled within the digital domain through a computational process that directly influences the real world manufacturing operations. The feedback received by the physical processes also influences computational procedures in an optimization feedback loop \cite{krupitzer2020survey}. Smart factories may consist of several CPPS whose role is not only to automate machine interaction, but to minimize production anomalies, thus maximizing efficiencies in production.

\noindent \underline{Collaborative Machines:}
Since the inception of cyber-physical production systems (CPPS), it is evident that the role of humans has been evolving to facilitate closer interaction between people and machines in ever-changing production environments. The concept of collaborative machines/robots \cite{salcic2017designing} in industrial automation started gaining momentum with the emergence of Industry 4.0 \cite{camarinha2017collaboration}, where these machines are assuming many of the responsibilities previously assigned to the human workforce \cite{bragancca2019brief}. However, it simultaneously poses a new set of challenges and requirements \cite{gualtieri2020opportunities} while dealing with these machines within a production facility. Context awareness, self organization, mutual control, and communication paradigms are some of the key aspects which need to be considered and which can be considered critical to enable synchronized operation in a collaborative industrial environment \cite{zhou2015industry}.

\noindent \underline{Human-Machine Interaction:}
While the emphasis of early scientific efforts in the domain of human-machine interactions (HMI) has been on fully controllable systems, it has quickly evolved to incorporate adaptable mechanisms involved with the development of highly complex and dynamic HMI systems \cite{gorecky2014human}. In such a complex HMI system, the human and machine agents can no longer be thought of in isolation, but rather as a collaborative unit accomplishing tasks in a distributed manner by assigning responsibilities in a distributed between participating entities \cite{nardo2020evolution}. Numerous studies conducted on HMI \cite{masood2019augmented, gattullo2019towards,fraga2018review, damiani2018augmented} primarily focus on the applications of Virtual Reality (VR) and Augmented Reality (AR) in a smart industry scenario.

\noindent \underline{Digital Twins:}
Digital Twins (DTs) enable industry to create digital copies of the physical products manufactured in an industry 4.0 situation. Industrial IoT digital twins optimize production efficiency by predicting failures in the production process. Such failures can then be mitigated for before they impact on manufacturing reducing factory down time and associated losses in revenue. Moreover, Digital Twins also enable remote commissioning and diagnostics for the off-the-shelf products to lower service costs and improving customer satisfaction. Some works \cite{tao2019digital, damjanovic2019open} in the literature throw light on the design and development of digital twin demonstrations for smart manufacturing through the use of open source technologies comprising software, hardware, or a combination of both. Furthermore, the authors also highlight the implementation requirements for CPPS and demonstrate on how CPPS and DTs can be used to  benefit  industry processes.        

\subsubsection{Video surveillance}
Artificial Intelligence (AI) based-Video Surveillance is one of the significant applications implmented in industry 4.0 which can significantly maximise effectiveness and production efficiency across the complete manufacturing environment \cite{thoben2017industrie}. The solutions typically available in this domain generally combine high resolution cameras, data storage abilities, and machine learning enabled management hardware to provide a smart industry with the insights to inform their ongoing operations \cite{reportvidsurv2018}. Moreover, video analytics makes it possible to detect any anomalous behavior on the production line and enables expert systems to trigger appropriate alerts on the basis of events captured through these surveillance cameras \cite{9055222sherazi2020}.

\begin{table*}[!htp]
 \centering
 \caption{Use cases for smart manufacturing along with their requirements \cite{8403588}.}
 \label{requirements}
 \makegapedcells
  \adjustbox{max height=\dimexpr\textheight-5.5cm\relax, max width=\textwidth}{
 \begin{tabular}{|*{9}{c|}}
  \toprule
  \multirowcell{5.2}[2.3ex]{\diagbox[height=4.3\line]{\rlap{\enspace\raisebox{2ex}{\textbf{Use Cases}}}}{\raisebox{-3.5ex}{\textbf{Requirements}}}}
  	&\multicolumn{5}{c|}{\textbf{Link Requirements}}
  	&\multicolumn{3}{c|}{\textbf{System Requirements}}\\
		\cline{2-6}
		&\multicolumn{3}{c|}{Time Critical}
		&\multicolumn{2}{c|}{Non Critical}
		&\multicolumn{3}{c|}{}\\
			\cline{2-9}
  &\makecell{Cycle Time} & \makecell{Payload} & \makecell{Jitter} & \makecell{Latency} & \makecell{Datarate} & \makecell{Service Area}& No. of Nodes&Mobility\\
  \hline
  Collaborative Machines & 4-10 \textit{ms} &\textless 1 \textit{KB} & \textless 1 $\mu$ s & \textless 10 \textit{ms} & \textless 5-10 \textit{Mbps} & 1 \textit{km} X 1 \textit{km} X 0.3 \textit{km} & 5-10 & - \\
  Human-Machine Interaction & \textless 10 \textit{ms}& 20-50 \textit{Mbit}& -& \textless 30 \textit{ms}& -& Typical factory floor size& \textless= no. of workers& Low\\
  Digital Twins& 50 \textit{ms}& 20-100 \textit{KB}& -& -& 5-10\textit{Mbps}& \textless 1 $km^{2}$ & 100& Fairly high\\
  
  Video Surveillance& 10-100 \textit{ms}& 15-150 \textit{KB}& \textless 50\% of cycle time& -& \textless 10 \textit{Mbps}& \textless 1 $km^{2}$ & 100& Fairly high\\
  Remote Operation& 50 \textit{ms} & Few \textit{Bytes} & \textless1-10 \textit{ms}& -& -& Several $km^{2}$ & \textless 10 \textit{k} & Low\\
  Predictive Maintenance& 50 \textit{ms} & - & - & - & - & Several $km^{2}$ & \textless 10 \textit{k}& Low\\
  \hline
 \end{tabular}
 }
\end{table*}

\subsubsection{Remote operation}
This refers to the remote monitoring and remote control of equipment and machinery in a remote production setting within a smart factory \cite{FRANK201915}. Remote operation enables the management of a smart industry to manage a range of operations remotely, from the procurement of raw material, manufacture of any products, to precisely meeting the regular order updates and ensuring a timely delivery of raw materials \cite{zhong2017intelligent}. Another example of remote operation is the field of  autonomous robotics. Autonomous robotic systems involve the remote operation of devices in real-time, for reasons that range from protecting human worker safety in hazardous work environments to applications that require greater speed efficiency or precision than human workers can deliver.

\subsubsection{Predictive maintenance}
There have been a significant number of organized research efforts in the diagnosis and prognosis of faulty mechanical systems in the past. However, the advent of the industrial revolution created a need for more smarter ways of looking after the machinery in a smart factory \cite{li2017intelligent, lee2019quality}. Predictive maintenance is an advanced condition monitoring based paradigm that uses smart tools and techniques to track the performance of in-service equipment to establish when the maintenance is required \cite{sezer2018industry} to avoid system breakdown before it happens.

\subsection{Use-case requirement analysis}
Recent years have witnessed significant efforts by  consortium  \cite{NGMNwhitepaper, 5GPPPwhitepaper} and industrial partners \cite{EUFP7Project, BMBFProject} to identify the requirements of smart manufacturing. 
For all the use cases discussed throughout this section, there are a set of communications and networking requirements to be met to enable smart manufacturing as summarized in Table \ref{requirements}. These requirements can broadly be defined in terms of link and system level requirements. The link requirements can further be classified into time critical (such as cycle time, payload, and Jitter) and non critical requirements such as  covering service area and number of nodes required.

\section{Existing Wireless Technologies In Smart Manufacturing}\label{sec:exis_tech}

A wide range of wireless networking technologies exists that have applicability in smart manufacturing. To ascertain a wireless technology’s suitability for a particular use case application, one needs to compare the various technologies' properties and capabilities. For example, let's consider the physical layer of different existing wireless standards such as IEEE 802.11 for wireless local area networks (WLAN), IEEE 802.15.1 for WPAN/Bluetooth, or IEEE 802.15.4 for low-rate wireless Private Area Network (PAN). These communications protocols operate in one or more ISM (Industrial, Scientific and Medical) frequency bands: 2.4 GHz, 5 GHz, and 868 MHz, and each has its pros and cons (e.g., increased coverage at lower frequencies, higher data rate at higher frequencies etc.). Unfortunately, there is still a real problem of overlapping of certain frequency bands which results in partially blocking of frequencies or interference.
\begin{table*}[h!]
\caption{Overview of Wireless Technologies}
\label{table1}
\begin{tabular}{|cc|c|c|c|}
\hline
\multicolumn{2}{|c|}{\textbf{Name}} &
  \textbf{\begin{tabular}[c]{@{}c@{}}Range \\ (m)\end{tabular}} &
  \textbf{\begin{tabular}[c]{@{}c@{}}Data Rate\\ (Mbps)\end{tabular}} &
  \textbf{Infrastructure needs} \\ \hline
\multicolumn{1}{|c|}{\multirow{4}{*}{\begin{tabular}[c]{@{}c@{}}Short to medium range \\ wireless technologies\\ (IEEE 802.15.1 )\end{tabular}}} &
  Bluetooth &
  100 &
  \textless 3Mbps &
  \begin{tabular}[c]{@{}c@{}}No special infrastructure, \\ point to point (P2P)\end{tabular} \\ \cline{2-5} 
\multicolumn{1}{|c|}{} &
  WISA &
  \multirow{3}{*}{5 - 15} &
  \multirow{3}{*}{1 Mbps} &
  \begin{tabular}[c]{@{}c@{}}Wireless proximity switch(s), \\ wireless sensor pad(s), \\ and wireless input/output pad(s)\end{tabular} \\ \cline{2-2} \cline{5-5} 
\multicolumn{1}{|c|}{} &
  WSAN-FW &
   &
   &
  WSAN-FA base station(s), and converter(s) \\ \cline{2-2} \cline{5-5} 
\multicolumn{1}{|c|}{} &
  IO-Link &
   &
   &
  \begin{tabular}[c]{@{}c@{}}IO-Link Wireless Device(s) (Hub) \\ and Wireless Bridge(s)\end{tabular} \\ \hline
\multicolumn{1}{|c|}{\multirow{5}{*}{\begin{tabular}[c]{@{}c@{}}Short to medium range \\ wireless technologies\\ (IEEE 802.15.4)\end{tabular}}} &
  Zigbee &
  15 &
  250 Kbps &
  Access points \\ \cline{2-5} 
\multicolumn{1}{|c|}{} &
  Wireless HART &
  15 &
  250 Kbps &
  HART-Gateway to the fieldbus \\ \cline{2-5} 
\multicolumn{1}{|c|}{} &
  ISA100.11a &
  15 &
  250 Kbps &
  Gateways and bus \\ \cline{2-5} 
\multicolumn{1}{|c|}{} &
  \begin{tabular}[c]{@{}c@{}}WIA-PA \\ (IEC62601)\end{tabular} &
  10-100 &
  250 Kbps &
  \begin{tabular}[c]{@{}c@{}}Host computer, gateway, \\ routing device(s), \\ and handheld device(s)\end{tabular} \\ \cline{2-5} 
\multicolumn{1}{|c|}{} &
  6TiSCH &
  15 &
  250 Kbps &
  Multiple Gateways and relays \\ \hline
\multicolumn{1}{|c|}{\multirow{3}{*}{\begin{tabular}[c]{@{}c@{}}Short to medium range \\ wireless technologies\\ (IEEE 802.11)\end{tabular}}} &
  WLAN &
  100 &
  600 Mbps &
  Router, access points \\ \cline{2-5} 
\multicolumn{1}{|c|}{} &
  Industrial WLAN &
  100 &
  450 Mbps &
  Access points, gateways to fieldbus \\ \cline{2-5} 
\multicolumn{1}{|c|}{} &
  \begin{tabular}[c]{@{}c@{}}WIA-FA\\ (IEC 62948)\end{tabular} &
  5–30 &
  \textless 54 Mbps &
  \begin{tabular}[c]{@{}c@{}}Host computer, gateway, access device(s), \\ field device(s), and handheld device(s)\end{tabular} \\ \hline
\multicolumn{1}{|c|}{\multirow{4}{*}{\begin{tabular}[c]{@{}c@{}}Short to medium range \\ wireless technologies\\ (others)\end{tabular}}} &
  UWB &
  10 &
  1 Gbps &
  Locator devices \\ \cline{2-5} 
\multicolumn{1}{|c|}{} &
  EnOcean &
  30 &
  125 Kbps &
  Transceiver modules \\ \cline{2-5} 
\multicolumn{1}{|c|}{} &
  NFC &
  10 cm &
  106 - 424 Kpbs &
  Transceiver modules \\ \cline{2-5} 
\multicolumn{1}{|c|}{} &
  RFID &
  6 &
  100 Kbps &
  Tags, scanner \\ \hline
\multicolumn{1}{|c|}{\multirow{4}{*}{\begin{tabular}[c]{@{}c@{}}Long range \\ wireless technologies\end{tabular}}} &
  LoRa/ LoRaWAN &
  \textless{}10 km &
  \textless{}= 27 Kbps &
  \begin{tabular}[c]{@{}c@{}}One or more gateways, \\ application server or \\ cloud system\end{tabular} \\ \cline{2-5} 
\multicolumn{1}{|c|}{} &
  NB-IoT &
  \begin{tabular}[c]{@{}c@{}}1 km (urban) \\ 10 km\\ /rural)\end{tabular} &
  \textless 100 Kbps &
  Infrastructure from provider \\ \cline{2-5} 
\multicolumn{1}{|c|}{} &
  LTE-M &
  \begin{tabular}[c]{@{}c@{}}500 m (urban)\\ 5 km (rural)\end{tabular} &
  0.384 – 1 Mbps &
  Infrastructure from provider \\ \cline{2-5} 
\multicolumn{1}{|c|}{} &
  SigFox &
  \begin{tabular}[c]{@{}c@{}}10 km \\ (urban)\\ 40 km \\ (rural)\end{tabular} &
  100 or 600 bps &
  Infrastructure from provider \\ \hline
\multicolumn{2}{|c|}{LTE} &
  10 km &
  150 Mbps &
  \begin{tabular}[c]{@{}c@{}}Complex infrastructure from \\ provider\end{tabular} \\ \hline
\multicolumn{2}{|c|}{WIFI 6} &
  10m &
  \begin{tabular}[c]{@{}c@{}}upto 11 Gbps \\ For 3 channels\end{tabular} &
  Router, access points \\ \hline
\end{tabular}
\end{table*}


Table~\ref{table1} gives an overview of the most relevant wireless technologies/solutions which can be used for various smart manufacturing applications. It highlights different technologies and their enabling features such as communication range, data rate etc., along with the infrastructure requirements of the corresponding wireless technology. Based on the feasible communication distance/range, existing wireless solutions can be broadly classified into two categories; short to medium range and long-range wireless technologies.

\subsection{Short to medium range industrial wireless technologies}\label{sec:Short_Range}  

This section addresses various short to medium range (up to 100m) wireless technologies which can be used for smart manufacturing applications. These technologies mainly fall under three different IEEE standards, namely, 802.15.1, 802.15.4 and 802.11:

\subsubsection{IEEE 802.15.1-based technologies} \label{sec:IEEE802}

\paragraph{Bluetooth}

Bluetooth is a short-range wireless technology based on the IEEE 802.15.1 standard which operates in the 2.4 GHz ISM frequency band. The initial versions of Bluetooth (versions 1 - 3), commonly referred to as classic Bluetooth, are widely used in cellular devices and personal computers \cite{Avnet_2020}. Even though it supported a higher data rate (1Mbps), classic Bluetooth was not widely adopted in industrial automation due to its high power consumption, short communication range, and lack of support for an increased number of nodes in a network. The introduction of ``version 4.0", known as ``Bluetooth Low Energy (BLE)", opened the market for this technology in the consumer product space as well as smart manufacturing applications. The key feature of BLE was its reduced power consumption and hence improved battery lifetime. BLE achieves low energy consumption by keeping the radio in standby mode for most of the time. It is observed that BLE consumes very little energy per bit transmitted when compared with other (e.g., IEEE 802.15.4 based ) wireless communication technologies \cite{siekkinen2012low}. The energy consumption of a few of the widely used BLE wireless devices are reported in \cite{lindh2017measuring}. The shorter communication range persists as a drawback of this technology in some applications due to its low power operation and the difficulty in scaling the network to cover a larger area. This problem of the reduced range and network size was solved with the recent introduction of Bluetooth Mesh Networking \cite{Mesh2018}, a low power personal area network (PAN) technology capable of supporting up to 32000 nodes in a many-to-many connection. This latest revision of the Bluetooth standard guarantees network reliability through a multi-path peer-to-peer connection. However, this standard is still very new, and no off-the-shelf solutions are available at the time of writing. However, its universality, distribution, and high data rate has the potential to make it one of the most widely used wireless protocols in industrial environments.  

\paragraph{WISA, WSAN-FA, IO-Link Wireless}

Wireless Interface for Sensors and Actuators (WISA) \cite{Steigmann2006} was initially developed by ABB, and the first WISA device became available in 2004. WISA is based on the IEEE802.15.1 physical layer (the same as standard Bluetooth). WISA based wireless systems are still in operation and demonstrate the viability of wireless networking in factory automation. One base station supports up to 120 wireless devices \cite{Steigmann2006}. WISA communication links connect sensors and actuators to an Input/Output module called the “base station”. At any particular time, a maximum of three base stations can be operated within one manufacturing cell “without significant loss of performance”. The communication range is 5~m for an industrial environment against a 15~m typical range. Latency time is typically 20~ms for 99.99\% of all cases, and 34 ms maximum \cite{Steigmann2006}. IEEE 802.11 (Wi-Fi) channels can be blacklisted in segments to improve coexistence with Wi-Fi deployments.

In 2010, ABB made the WISA technology specification available to the PROFIBUS and PROFINET user organization (PNO) to develop it into an open standard, called “Wireless Sensor Actuator Network for Factory Automation” (WSAN-FA) \cite{Rentschler2017}. The PROFIBUS and PROFINET user organization (PNO) published this technology as an industry-standard in 2012 \cite{PNO2012}. WSAN-FA also utilizes the physical layer of Bluetooth (IEEE 802.15.1) and provides synchronization between nodes by Frequency Hopping Multiple Access, a TDMA and Frequency Hopping combination. PNO designed this system for factory automation needs at the sensor and actuator level and used the IO-Link interface, and system specification standard \cite{IOLink2019} as the data format to integrate with other IO-Link devices and services. However, the WSAN-FA standard was never adopted by industry due to significant gaps in the specification \cite{Heynicke2018}.

\begin{figure}
    \centering
    \includegraphics[scale=0.5]{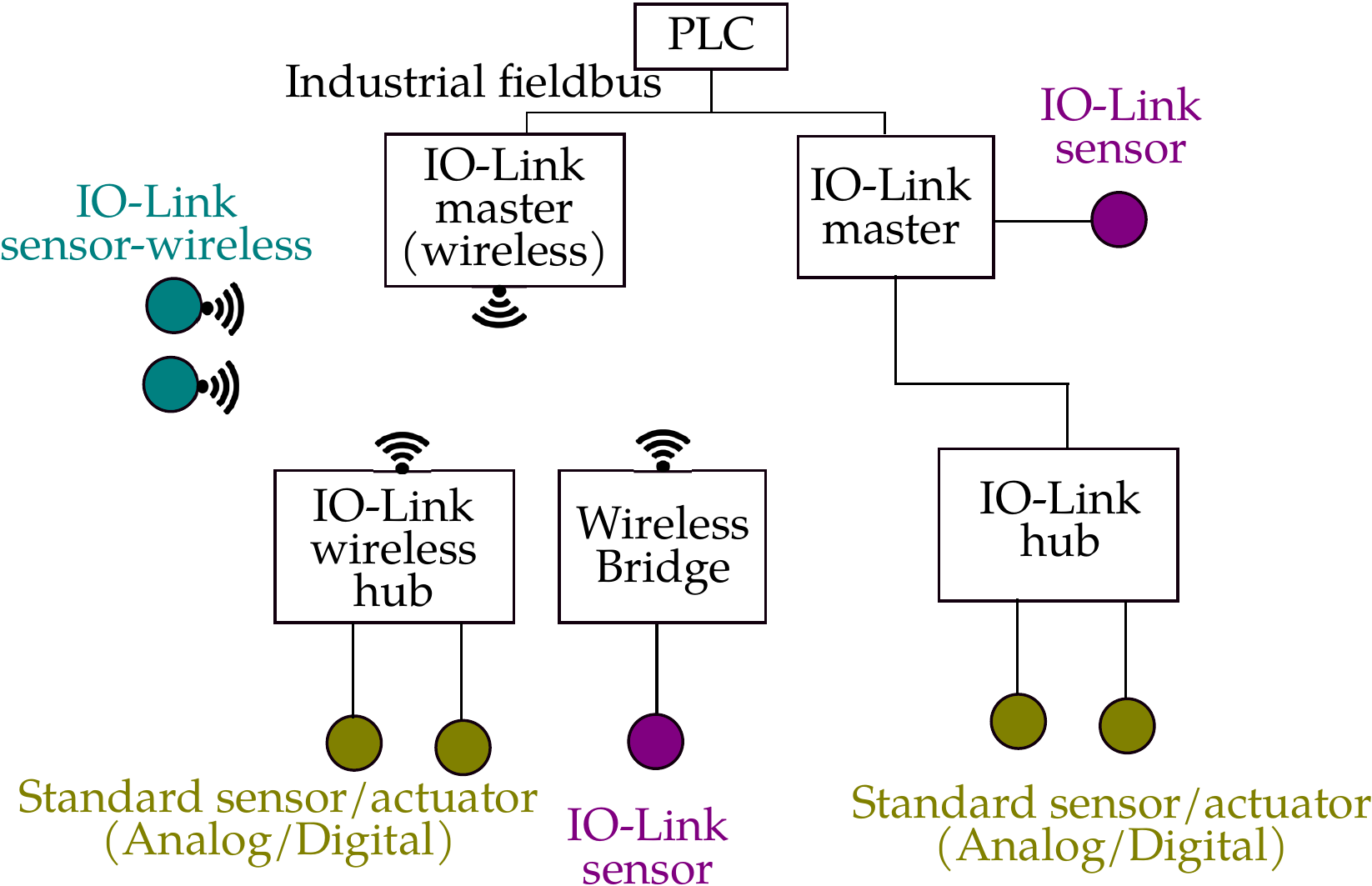}
    \caption{IO-Link system architecture \cite{Heynicke2018}}
    \label{fig_iolink}
\end{figure}

To address these gaps, the PNO and IO-Link consortium has developed the standard further into the IO-Link Wireless Systems Extensions (IOLW) \cite{IOLink2018}, which was published in 2019. Fig.~\ref{fig_iolink} shows IO-Link architecture, both wired and wireless. The IO-Link standard (IEC 61131-9, 2013) specifies a half-duplex, fieldbus-neutral,  point to point communication mechanism between the sensor/actuator and the controlling device. IO-Link architecture consists of a gateway (known as IO-Link master) having multiple ports, to which individual IO-Link devices are connected. These individual devices can be any I/O device such as sensors, actuators, RFID readers etc. The IO-Link system also contains the tools required for the configuration of sensors/actuators and parameter assignment.  From a users perspective there is not much difference between wired and wireless IO-Link systems except the fact that the individual devices get connected to the master wirelessly. As shown in Fig.~\ref{fig_iolink}, IO-Link wireless bridge can be used to connect the conventional IO-Link devices to a wireless master. IOLW supports roaming of wireless devices between access points. The hand-over mechanism of a roaming device is guaranteed  to be below 1s \cite{Heynicke2018}. IOLW works in the 2.4GHz ISM band as defined in 802.15.1. The cycle time between a master and a device can be optimized to 5 ms (allowing two re-transmits) \cite{Heynicke2018}.

IEEE 802.15.1 based technologies are currently used in industrial plants such as chemical, oil-gas, water, and power, as a part of monitoring and maintenance solutions. Ease of deployment, wide availability, low-power and low cost makes Bluetooth a superior choice for sensing and device monitoring, which falls primarily under the predictive maintenance use case category discussed in Section~\ref{sec:use_cases}. One example is in a motor assembly line that spans across $100~m \times 500~m$ factory floor at Mercedes-Benz Ludwigsfelde GmbH (Germany), Bluetooth is used to connect cordless power tools to a central programmable logic controller (PLC) \cite{Baumann_2020}.  Other experimental works are described in the research literature, which presents the applicability of Bluetooth technology in the smart industry \cite{Gore_2019}\cite{kadechkar_2020}. For example, in \cite{Gore_2019}, a BLE based autonomous sensor monitoring system for industrial plants and its performance evaluation is presented. A Bluetooth based real-time monitoring of a high-voltage substation connector is discussed in \cite{kadechkar_2020}. Nowadays, Bluetooth based industrial wireless products/solutions are available from several vendors in the market, such as the Bluetooth wireless modules from Phoenix Contact \cite{phoenixcontact}, the IO-Link devices from Baumer \cite{baumer} etc.  

\subsubsection{IEEE 802.15.4-based technologies}

IEEE 802.15.4 based wireless networks are another class of short-range, low-cost, low-power communication technology targeting wireless sensor networks. The basic IEEE 802.15.4 MAC schemes were deficient in supporting identified industrial application requirements such as reliability and real-time capability. The addressed improvements resulted in the evolution of various industrial wireless standards such as WirelessHART, ISA100.11a, WIA-PA (Wireless network for Industrial Automation – Process Automation), which are widely meant for process automation industries. In this section, we provide a brief overview of different industrial wireless solutions that are based on the IEEE 802.15.4 standard. 

\begin{figure*}[h]
\centering
\subfloat[WirelessHART]{\includegraphics[scale=0.5]{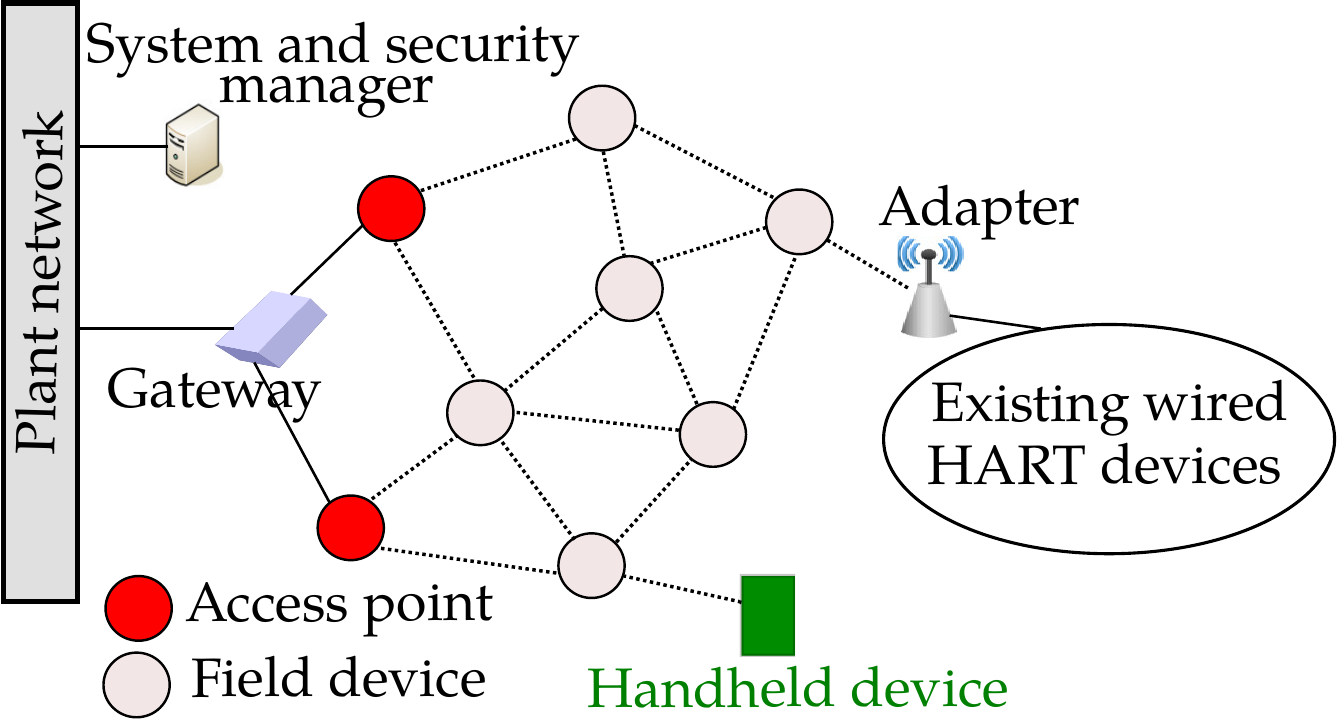}\label{Fig_WirelessHART}}
\hfil
\subfloat[ISA100.11a]{\includegraphics[scale=0.5]{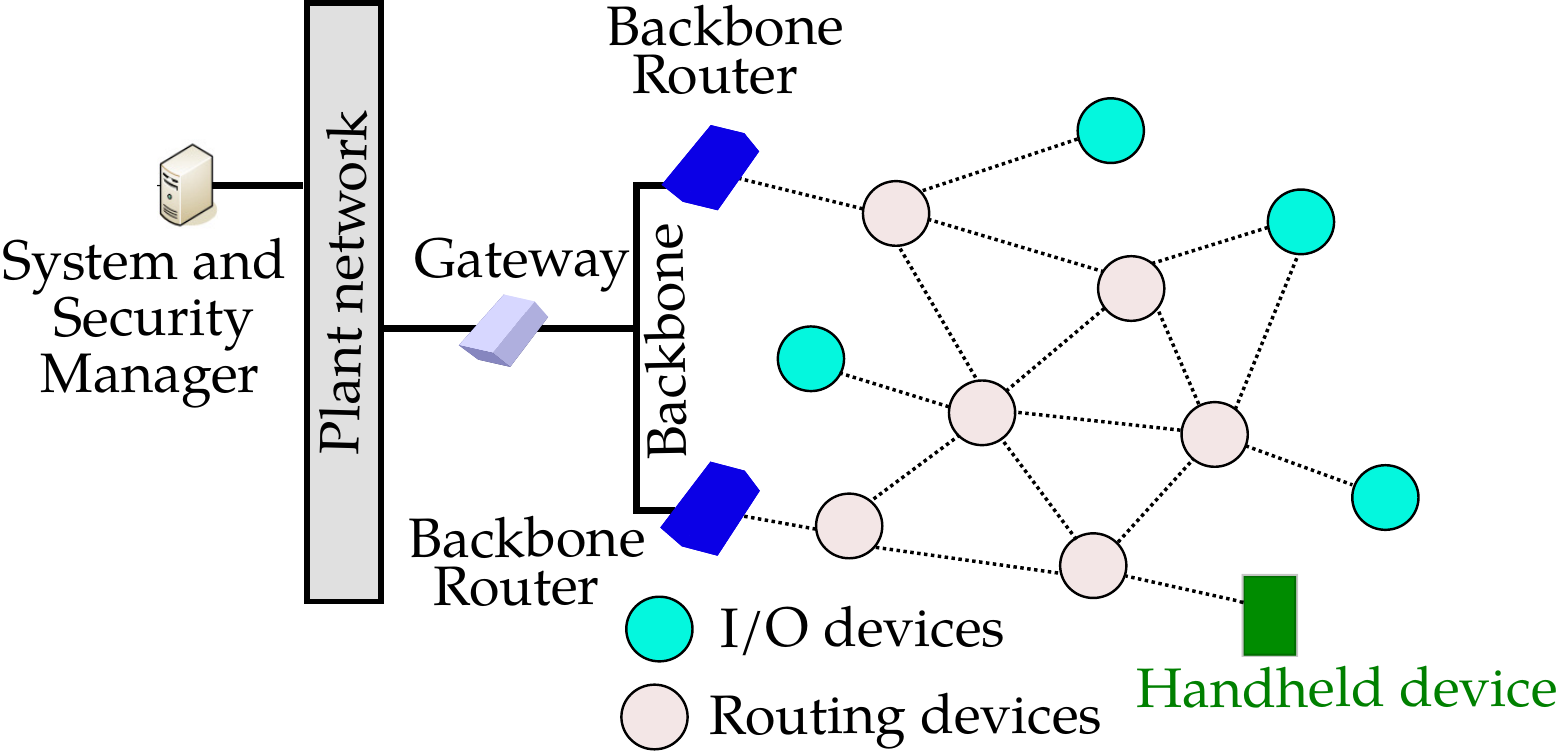}\label{fig_ISA100_11}}
\hfil
\subfloat[WIA-PA]{\includegraphics[scale=0.5]{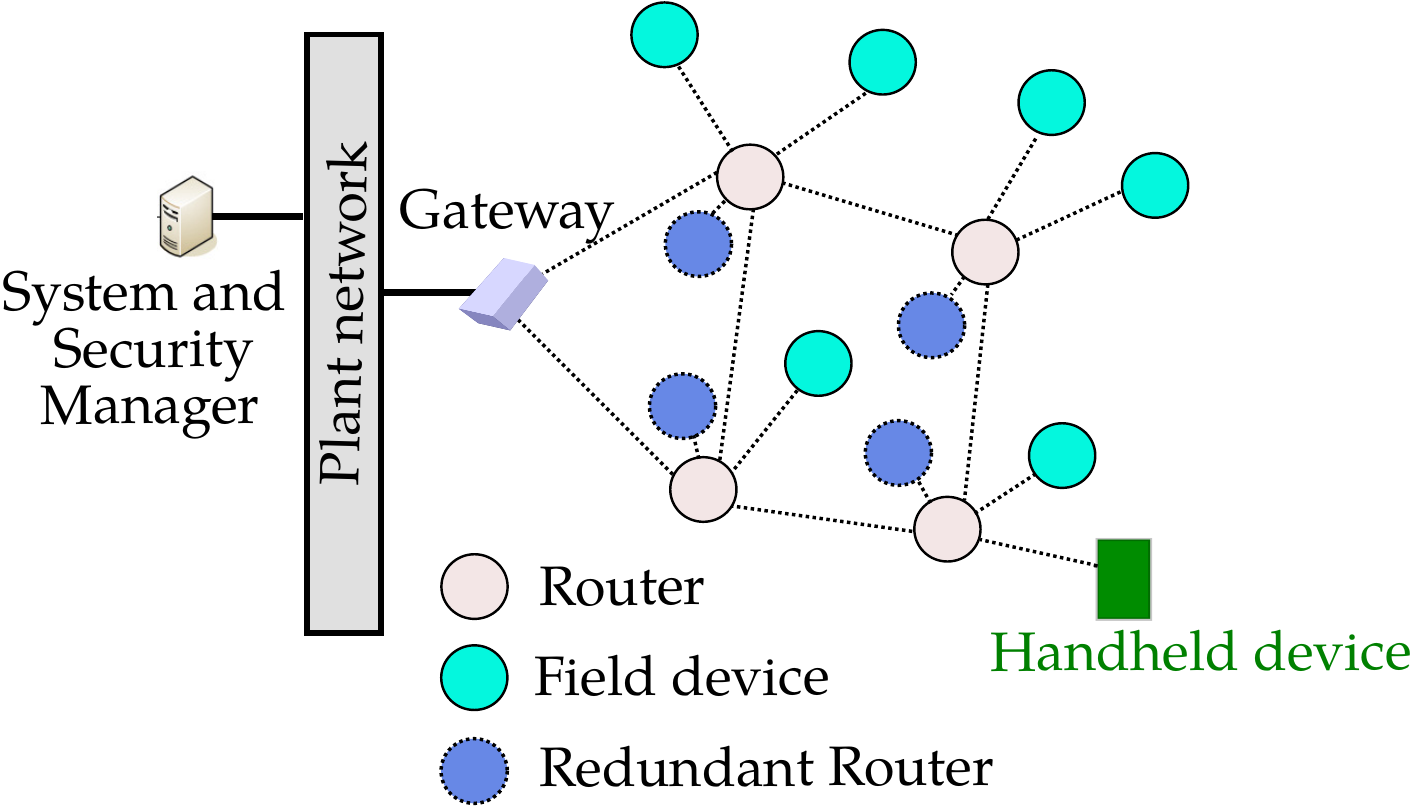}\label{fig_WIA_PA}}
\hfil
\subfloat[6TiSCH (Basic configuration)]{\includegraphics[scale=0.5]{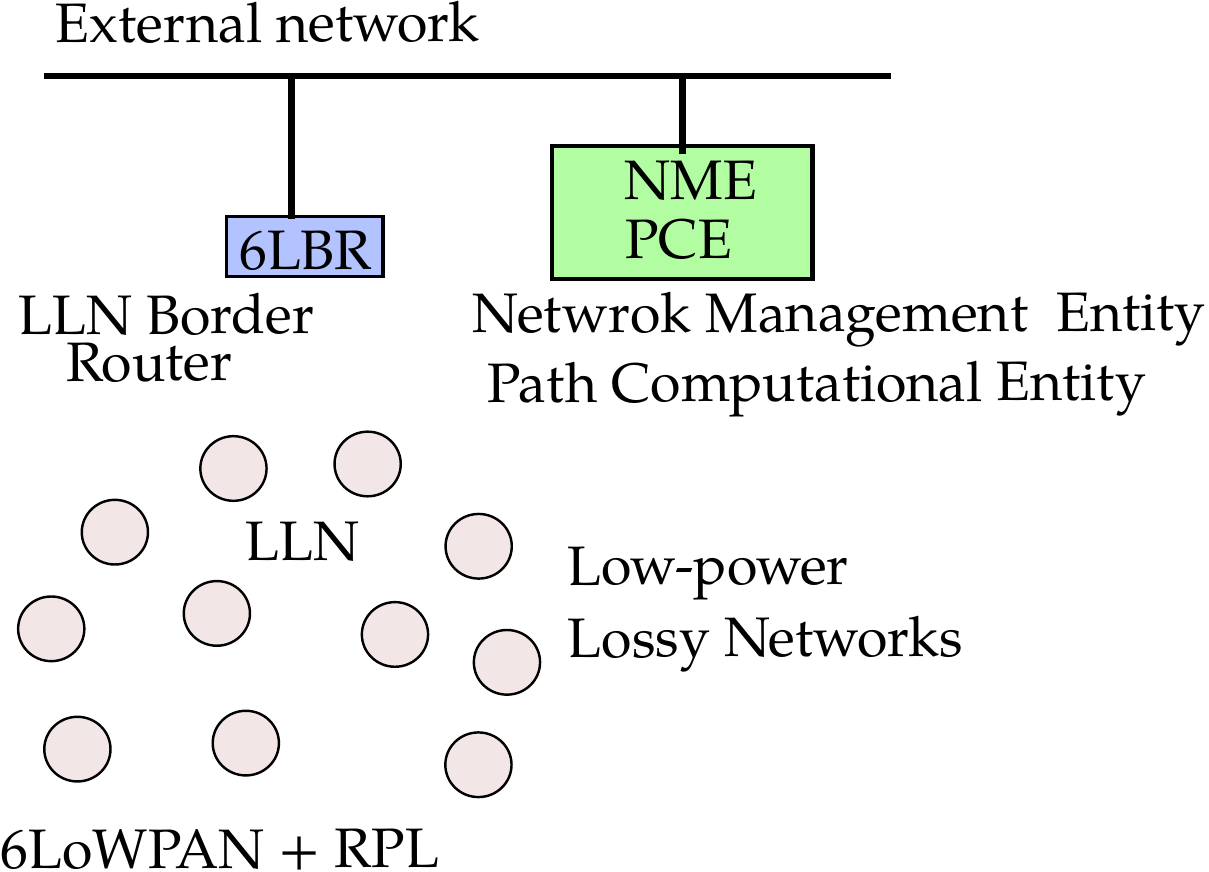}\label{fig_6TiSCH}}
\hfil
\label{datacollection_schemes_wsn}
\caption{IEEE 802.15.4 based industrial wireless solutions \cite{Petersen2011,wang2016comparative,ietf-6tisch}}
\end{figure*}

\paragraph{WirelessHART} 

WirelessHART \cite{Petersen2011} was specifically designed for process automation applications, based on an industrial automation protocol, HART (Highway Addressable Remote Transducer). At the physical layer (PHY), WirelessHART adopts IEEE208.15.4 PHY, and in the upper layers of the  communication stack it defines its own TDMA-based MAC layer \cite{Lennvall2008}. In this MAC layer, a network-wide time synchronization protocol is used, which is designed for unconfigurable 10ms time slots \cite{Saifullah2010}. In addition, WirelessHART introduces channel hopping and channel blacklisting into the MAC layer to deal with noise and interference in industrial environments \cite{Kim2010}. WirelessHART shares the same transport and application layers with the HART network stack.

A WirelessHART network consists of a set of field devices, adaptor(s), access point(s), gateway(s), a network manager, and a security manager, as shown in Fig.~\ref{Fig_WirelessHART}. Typically, 80-100 field devices are connected to one gateway in a star or mesh topology with the help of access points \cite{Saifullah2011,wang2016comparative}. Adaptors enable wireless functionalities to legacy HART field devices. The network manager provides network initialization functions, network scheduling, link routing, etc. Routing is performed through either graph routing or source routing \cite{Saifullah2015}.
The WirelessHART standard provides data confidentiality, data integrity, and authentication by using MIC \cite{Palattella2013}, CRC \cite{Lin2007}, and 128-bit AES keys \cite{Raza2009}. Different symmetric security keys (join, network, session, well-known) are used to secure end-to-end, per-hop, and peer-to-peer communication. The security manager, together with the network manager, is the responsible entity for establishing and distributing the security keys. 

\paragraph{ISA100.11a}

ISA100.11a is another industrial standard that runs on top of the IEEE 802.15.4 PHY (Physical layer) and exhibits many similarities with the WirelessHART. It also uses time-slotted communications and a slow channel hopping mechanism to deal with external interference. In ISA100.11a, a set of roles are defined to describe the functions and capabilities of a device, similarly to WirelessHART \cite{Petersen2011}.

ISA100.11a \cite{Dinh2012} was designed for process control in industrial automation and related applications \cite{Dinh2012}. ISA100.11a is built on the IEEE802.15.4 PHY layer and adopts many features in common with WirelessHART in the MAC layer, such as TDMA/CSMA, channel hopping, and channel blacklisting. However, ISA100.11a provides additional network flexibility over WirelessHART, such as configurable time slot, fast and slow channel hopping, and adaptive channel blacklisting \cite{Petersen2011}. The network and transport layers of ISA100.11a are based on 6LoWPAN, IPv6, and UDP standards \cite{Shelby2011}, \cite{Bormann2009}. In fact, ISA100.11a targets a wider class of process control applications than WirelessHART because ISA100.11a’s application layer is not restricted to one native protocol.

As shown in Fig.~\ref{fig_ISA100_11}, a typical ISA100.11a network consists of a set of field devices (I/O devices and routing devices) and infrastructure devices (backbone routers, gateways, and system and security manager). ISA100.11a operates in either tree or mesh topology while the second is preferable because it offers increased robustness and enhanced reliability \cite{Rezha2013}. Like WirelessHART, ISA100.11a supports graph routing and source routing \cite{Quang2013}. 
A security manager is essential in ISA100.11a, as in WIrelessHART, to distribute and manage the security keys for both message integrity check (MIC) as well as confidentiality. Five symmetric security keys (global, join, master, datalink, and session) are defined in ISA100.11a for encryption at different levels of the network stack. Optionally, ISA100.11a also defines asymmetric keys \cite{Zhang2009}.

Although there are similarities between ISA100.11a and WirelessHART, they are not interoperable \cite{Nixon2012}. From 2010 till 2013, ISA100.12 spent a lot of effort to find a technical path to converge with WirelessHART without success \cite{Song2008}.

\paragraph{WIA-PA (IEC62601)} 

WIA-PA also was designed to target process automation applications in industry setups. WIA-PA, like WirelessHART and ISA100.11a, builds upon the IEEE 802.15.4 standard. However, unlike these two technologies, WIA-PA avoided any modification to the IEEE 802.15.4’s MAC layer \cite{Liang2011}. This approach assures seamlessly co-existence with IEEE 802.15.4 MAC-based systems such as ZigBee. In order to ensure reliability and timeliness, WIA-PA defined a datalink sublayer on top of the IEEE 802.15.4 MAC layer. This sublayer provides additional functionalities like frequency switching, adaptive frequency hopping, packet aggregation \& disaggregation, and time synchronization, etc \cite{Zheng2017}.

WIA-PA supports a hierarchical network topology that is basically mesh-of-stars as shown in Fig.~\ref{fig_WIA_PA}. The first level of the network is a mesh topology of routers and gateways while the second level is a star network that connects redundant routers and field devices to routers from the first network level. A static routing method is used within a WIA-PA network \cite{Zhong2010}. The network manager sets up the connection relationships. Each pair of devices will have at least two routing paths between  \cite{Zhong2010_2}.
Security services such as data integrity (MIC) and confidentiality (encryption) are supported in WIA-PA at two levels: end-to-end and point-to-point. WIA-PA defines three symmetric keys (join, encryption, and data encryption) for encryption \cite{Min2010}, \cite{Wei2011}.

\paragraph{6TiSCH (IEEE 802.15.4-TSCH)}

6TiSCH (IPv6 over the TSCH mode of IEEE 802.15.4e) is a wireless standard meant to provide IPV6 connectivity for low power wireless networks consisting of IEEE 802.15.4 devices. It mainly makes use of the Time Slotted Channel Hopping (TSCH) mechanism detailed in IEEE 802.15.4-2015 standard. The basic architecture of a 6TiSCH network is shown in Fig.~\ref{fig_6TiSCH}. The architecture consists of a device called a ``Border Router", which acts as the intermediate gateway between the wireless network composed of low power devices and the external world (internet). 6TiSCH stack uses Constrained Application Protocol (CoAP), User Datagram Protocol (UDP) and IPV6 as the application, transport, and network-level protocols \cite{Vilajosana2019}. IPv6 Routing Protocol for Low-Power and Lossy Networks (RPL) is used to construct a tree-based multi-hop architecture to connect all the low power wireless nodes into the network. 6LoWPAN (IPv6-based Low-Power Personal Area Networks) adaptation layer handles the fragmentation/reassembly of IPV6 packets to IEEE 820.15.4 MAC packets and vice versa. 6TiSCH has also defined secure, lightweight join processes through Constrained Join Protocol (CoJP) \cite{Vilajosana2019}.

The communication between each pair of nodes in an IEEE 80.2.15.4 TSCH network follows a synchronous schedule consisting of ``dedicated cells". Each dedicated cell consists of a dedicated timeslot and a communication channel, with which a particular communication takes place between the respective nodes. This cell allotment scheme helps to avoid collisions and hence improve reliability. 6TiSCH supports both central and local management of these communication schedules using various scheduling functions. The allocation can be done such that the predictable transmission pattern matches the traffic. This avoids idle listening and extends battery life for the constrained nodes. 

Applications in process automation industries mainly involve monitoring and control of different fluids and their requirements can be broadly classified into three categories, namely monitoring, control and safety \cite{wang2016comparative}. The time critical applications related to safety and closed-loop control are still handled using wired communication technologies. Currently IEEE 802.15.4 based wireless solutions (WirelessHART, ISA100.11a, WIA-PA) are primarily employed for non-time critical applications such as monitoring and open loop control. The recent IEEE 802.15.4 architecture 6TiSCH, is expected to address mission-critical machine-to-machine communication. Industrial automation control systems, asset tracking with mobile scenarios, drones and edge robotic control are a few of the envisaged applications of 6TiSCH. Thus we believe that the IEEE 802.15.4 solutions can play a vital part in addressing some of the smart manufacturing applications which falls under different use cases identified in Section~\ref{sec:use_cases} such as, collaborative machines, remote operation and predictive maintenance.

\subsubsection{IEEE 802.11-based Technologies} \label{sec:IEEE802_11} 
IEEE 802.11-based wireless technology is well known by the common name ``Wi-Fi" (Wireless Fidelity) and provides wireless connectivity to the majority of today's digital devices such as mobile phones, laptops, smart TVs etc. The most commonly used wireless technologies in smart manufacturing, especially discrete manufacturing industries are also IEEE 802.11 based. In this section we provide a brief overview of IEEE 802.11 and its applicability in industrial automation and smart manufacturing domains. 

\paragraph{Wi-Fi Variants}

Wireless fidelity (Wi-Fi) is based on IEEE 802.11 (a/b/g/i/e/n/ac/ax) standards for wireless local area networks (WLAN). The main purpose is to provide wireless connectivity for devices within a local area. It provides an internet connection when connected to an Access Point (AP) and also supports station mobility \cite{Lee2007}.It standardises access to frequency bands for the purpose of local area communication. It also defines the MAC and physical layer specifications for wireless connectivity for fixed, portable, and moving stations (STAs) \cite{817038}. Wi-Fi based implementations for wireless sensor networks (WSN) have been analyzed in \cite{5975693}, and it has been pointed out how Wi-Fi can be more advantageous with respect to traditional WSN due to:
\begin{itemize}
\item	Higher data rate (up to 300 Mbps), useful for real time applications;
\item	Non-line-of-sight transmission: communication can also take place through walls;
\item	Large coverage area (100 m indoor, 300 m outdoor);
\item	High reliability to treat the network handling and fault recovery.
\end{itemize}

Although Wi-Fi consumes more power than other standards (e.g. IEEE 802.15.1 and IEEE 802.15.4), Wi-Fi is suitable for real time and high data rate implementations (e.g., audio/video surveillance) because of its low energy per bit rate (mJ/Mbit) \cite{Lee2007_2}.

\begin{figure}
    \centering
    \includegraphics[scale=0.7]{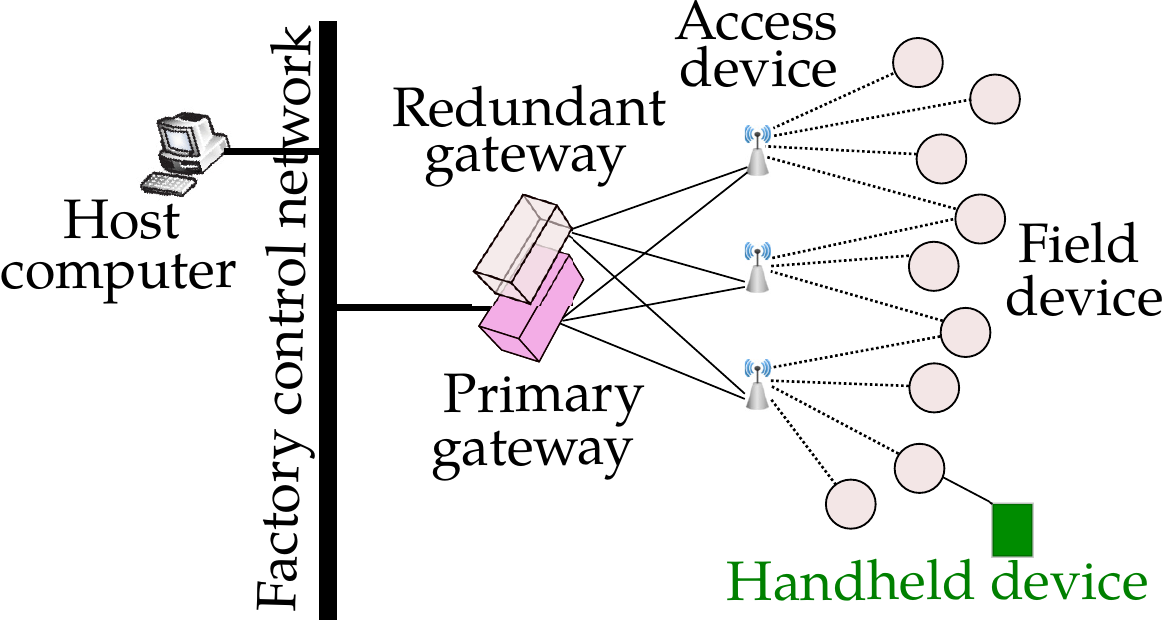}
    \caption{WIA-FA architecture (Redundant star topology) \cite{Liang_2019}}
    \label{fig_WIA_FA}
\end{figure}

\paragraph{WIA-FA (IEC 62948)}

Wireless Networks for Industrial Automation – Factory Automation (WIA-FA) has recently become an international standard, IEC 62948 \cite{IEC62948} and the system architecture is based on the physical layer of Wi-Fi (IEEE 802.11 WLAN). Therefore, the RF bandwidth is at least 20 MHz, which means it needs to be carefully installed, as a maximum of three parallel Wi-Fi based systems can operate in the 2.4GHz band without interference or degradation of performance. 

Fig.~\ref{fig_WIA_FA} shows the system architecture of WIA-FA.
Field devices, e.g. sensors or actuators, connect to an access device. Multiple access devices may communicate with the field devices in parallel and form multiple, optionally redundant, star topologies. Access devices have the same address and are transparent to the field devices. One or several access devices connect to one of a number of typically redundant gateway devices, which include the network management functions and connect a WIA-FA network to other networks within a factory environment. 

Each WIA-FA network has only one network manager (NM), which resides in the gateway device where it implements the network management function. The WIA-FA network management protocol performs such functions as allocating the unique 8-bit or 16-bit short address for all devices in the network; constructing and maintaining the redundant star topology; allocating communication resources for communications of WIA-FA devices; and monitoring the status of the WIA-FA network, including device status and channel condition.

WIA-FA defines the physical layer (PHY), data link layer (DLL) and application layer (AL). The PHY layer is based on the IEEE 802.11-2012 standard. DLL layer is designed in such a way that it provides  reliable, real-time and secured communication between WIA-FA field devices and access devices by adopting time-division multiple access (TDMA) data-transport mechanism based on the ``superframe" concept. Superframes are used to avoid transmission collisions between frames and ensure reliability and real-time transmission while supporting frame aggregation/disaggregation.  The WIA-FA superframe is a collection of timeslots which repeats at a constant rate. Though the length of a timeslot is configurable, each timeslot is only used for transmitting one frame. The default superframe consists of beacon timeslots (used by a field device to join the network), uplink shared timeslots, and downlink timeslots. The DLL is also responsible for management functions, including defining device joining, leaving, time synchronization, and remote attribute get/set.

The WIA-FA application layer (AL) supports distributed applications for users. The AL is comprised of the user application processes (UAP) and the application sub-layer (ASL), which defines communication services among UAPs on different devices. Each UAP is composed of one or more user application objects (UAO) that interact with industrial processes, while the device management application process (DMAP) is a special UAP.
WIA-FA supports a number of application data types that are transferred between the gateway device and field devices:
\begin{itemize}	
\item Non-periodic urgent commands such as start and stop commands with data priority RT0; 
\item	Periodical input data (e.g., sensor measurement values, switch status, actuator feedback values), and periodical output data (e.g., actuator setpoints, switch set values) with data priority RT1;
\item	Non-periodical requests and responses for attribute read-and-write accesses, as well as alarm acknowledgements with data priority NRT (non-real-time); 
\item	Non-periodical alarm reports with data priority RT2; and,
\item	Periodic management data (priority RT3) for monitoring data and network status messages.
\end{itemize}

\subsection{Long Range Wireless Technologies} \label{sec:Medium_Range} 

\subsubsection{LoRa and LoRaWAN} 
\begin{figure}[h]
\centering{\includegraphics[width=220pt]{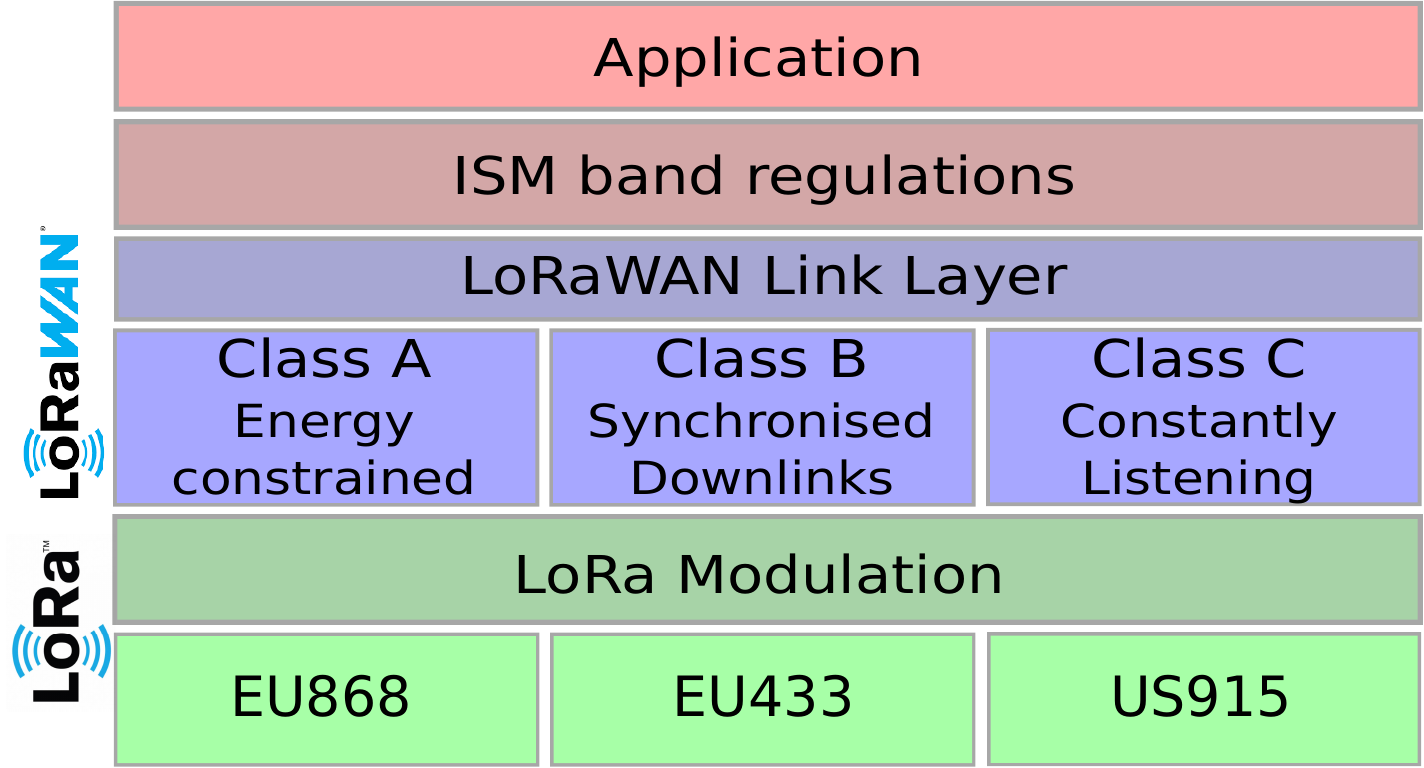}}
\caption{LoRaWAN classes and license-free sub-gigahertz radio frequency bands for the EU and the US}\label{figure1}
\end{figure}

LoRa (Long Range) is a proprietary radio technology with long range capabilities as well as high resistance to interference. Thanks to the spread-spectrum modulation technique LoRa can trade data rate with sensitivity. The amount of spread used is controlled through a parameter called the Spreading Factor (SF). The higher the SF, the higher the sensitivity and thus, the longer the transmission range. However, the data rate decreases substantially with higher SFs with a corresponding increase in energy consumption. Other characteristics that can affect the transmission time and the energy consumption is the channel bandwidth and the coding rate. A lower coding rate value results in shorter transmission times but less tolerance to transmission errors. LoRa can be used in both license-free sub-GHz and 2.4GHz radio frequency bands. Depending on the region the central frequency may be the 433MHz (Europe and Asia), the 868MHz (Europe) or the 915 MHz (Australia and North America) as is depicted in  Figure~\ref{figure1}. 

The current LoRa-based standard for sub-GHz transmissions is called LoRaWAN. This protocol has been developed and is maintained by a non-profit association called the LoRa Alliance. LoRaWAN provides a number of services for LoRa-enabled devices such as device registration, acknowledgements, and end-to-end security. 


Unlike some short/medium range technologies such as those based on the IEEE 802.15.4 radios, which use multi-hop deployments, LoRaWAN uses 1-hop deployments in a star topology where the nodes communicate with one or more gateways and the gateways forward their packets to a network server via a backhaul network. 
Due to its Aloha-based MAC and the radio duty cycle restrictions, LoRaWAN as currently standardised, cannot guarantee packet delivery and low latency, two requirements of many industrial applications. However, it is suitable for applications that generate small amounts of data (e.g., predictive maintenance) and for applications that require high levels of mobility (e.g., asset tracking).

\subsubsection{Sigfox}
Sigfox is an inexpensive, reliable, low-power wireless protocol solution that connects sensors and devices, which is a part of the Low-Power Wide Area Networks (LPWAN) suite of technologies, deployed mainly for the development of the IoT networks. Sigfox  can provide two-way, secured communication services to a range of services within the IoT ecosystem. The modulation technique utilised by Sigfox is ultra-narrow band, which provides a security advantage: it has the ability to continue operating despite jamming attempts. Ultra-narrow band modulation has intrinsic ruggedness as the signal overlapping with the noise is very low.

Sigfox is designed for the transmission of small payloads over long distances and has a transmission range when transmitting these small payloads up to a range of 40km \cite{8660398}. Sigfox utilizes bandwidth very efficiently and experiences very low noise levels, resulting in high receiver sensitivity, ultra-low-power consumption, and inexpensive antenna design making it a suitable protocol for IoT use cases. All these benefits come at an expense of a maximum throughput of only 100 bps. The achieved data rate clearly falls at the lower end of the throughput offered by most other LPWAN technologies limiting application opportunities for Sigfox.

Sigfox operates in Europe using the public 868MHZ to 868.2 MHz band, with a bandwidth of 192KHz. Whilst in the rest of the world, Sigfox uses the bands between 902 and 928 MHz with restrictions according to local regulations. Each message is 100 Hz wide and transferred with a data rate of 100 or 600 bits per second \cite{Zuniga2016}.

The successful deployment of Sigfox within a manufacturing environment is very dependent on the specific use cases. For example, if a system is needed to transmit minimal data such as temperature every hour or oil-tank levels on a daily basis for example, then the deployment of Sigfox is a viable option. Another potential use case is in supply chain management, a Sigfox sensor on a pallet of important goods can be tracked across all the continent of Europe relaying e.g. temperature and location allowing the level of traceability that is now relatively inexpensive \cite{8660398, Zuniga2016, Mekki2018, 7945893}. 

\subsubsection{NB-IoT and LTE-M}
Narrow Band IoT (NB-IoT) is a 3GPP open standard for low-power and low data rate devices such as sensors and actuators. In contrast with the other two major players in LPWAN (i.e., LoRa and Sigfox), NB-IoT is designed to operate over cellular licensed radio frequencies. NB-IoT uses the existing 2G-4G network infrastructure and its transmissions (packets) are accommodated within a subset of the existing frame layout provided by LTE (4G). Consequently, NB-IoT uses a Frequency Division Multiple Access (FDMA) policy to access the medium for uplinks and Orthogonal FDMA (OFDMA) for downlinks. The modulation is the quadrature phase-shift keying (QPSK) over a 200kHz-width channel (GSM) or a 180kHz-width channel (LTE). Unlike LoRa and Sigfox, NB-IoT can achieve much higher data rates of 200 kbps for downlinks and 20 kbps for uplinks. The maximum allowed payload size is also much higher than the other two major LPWAN players (i.e., 1600 bytes).

As is stated in \cite{ratasuk2016nb}, the strengths of NB-IoT are (a) the high levels of coverage possible while maintaining a data rate of at least 160 bps, (b) the connectivity of a massive number of devices (52547 devices per cell), (c) low implementation complexity for IoT applications, and (d) low latency ($<$10 seconds for 99\% of the devices). A negative point of NB-IoT is the increased overhead due to the synchronisation mechanism. This leads to a battery lifetime of less than 3 years for applications with data periodicity of 2 hours \cite{ratasuk2016nb}. Compared to the other two major players in LPWAN, NB-IoT seems to achieve more reliable communications \cite{mroue2018mac} due to its synchronized nature and much higher data rates \cite{sinha2017survey}. A disadvantage of NB-IoT is that it relies on existing infrastructure and it does not give the flexibility to the users to deploy their own private network. This must be done through the infrastructure of the network operator.

\section{Emerging Wireless Technologies for Industrial Automation} \label{sec:Emerging_Tech}

The evolution of Industry 4.0 envisions use cases that have communication link and system requirements that may not be fulfilled by existing technologies.
The increasing demand for higher throughput, lower latency and higher user capacity in smart manufacturing applications led the demand for more capable technologies.
In this section, we provide an overview of the emerging wireless technologies that are being considered, their new features, and how they support the new use cases associated with Smart Manufacturing. 

\subsection{5G} 
The 5G cellular network aims at highly adaptable, converged and pervasive wireless information exchanges, and is expected to be a game changer, unlocking novel opportunities, services, applications and a wide range of use cases. In terms of wireless networking, 5G is intended to allow three distinct types of services namely, enhanced mobile broadband (eMBB), massive machine-type communications (mMTC) and ultra-reliable low-latency communications (URLLC). The eMBB aims to have uplink and downlink data speeds up to 10 Gbps and 20 Gbps, respectively.  The mMTC will enable autonomous environments in industry by allowing interaction between a large number (e.g., up to 1 million devices per square km) of smart sensors and gateways. For single transmission, URLLC aims at a 1 ms over-the-air round-trip time (RTT) and 99.999\% reliability, which is important for time-critical control applications. The correlation between the above services and smart manufacturing use cases is depicted in Fig.~\ref{fig:5g_triangle}.
To support these three types of services, 3GPP has introduced a unified air interface, namely 5G New Radio (NR), that can flexibly address the requirements of each service along with the following major features:

\begin{figure}[b]
    \centering
    \includegraphics[width=\linewidth]{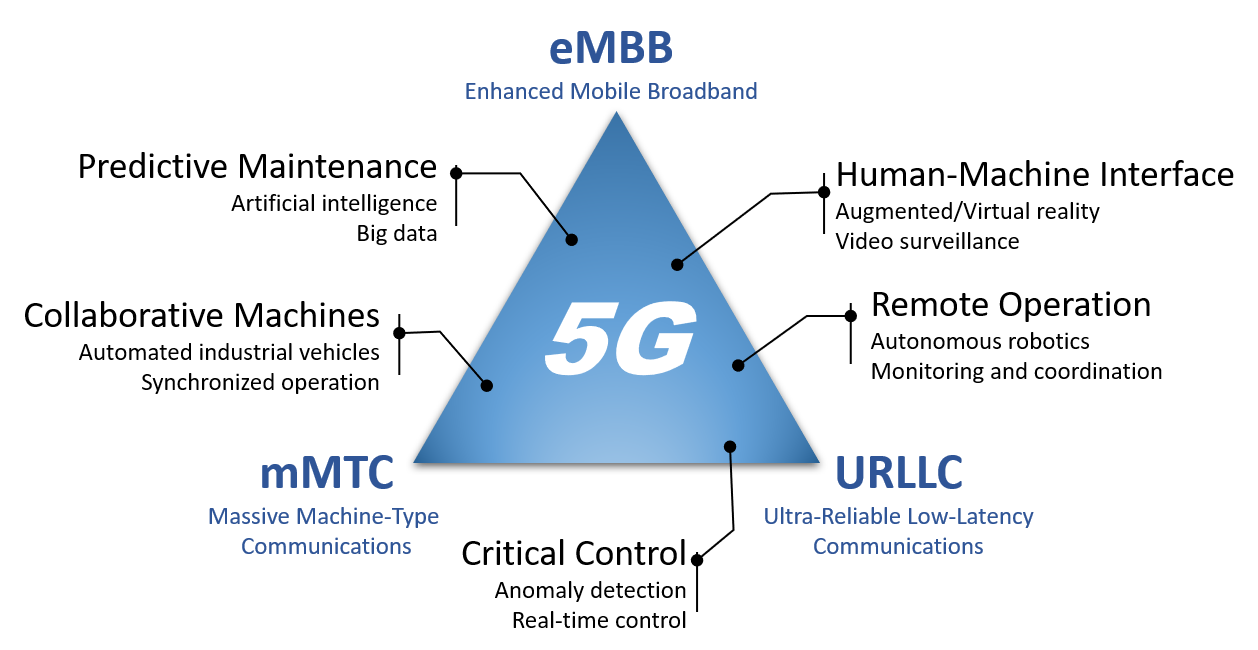}
    \caption{5G categories for smart manufacturing use cases.}
    \label{fig:5g_triangle}
\end{figure}
\begin{itemize}

\item {\bf New Radio (NR) Interface:} 3GPP has introduced a unified air interface, namely 5G New Radio (NR) that addresses many of the 5G services requirements and supports a large number of 5G use cases. The new radio interface/access will surpass the capabilities of previous generations of mobile communication. Due to the introduction of flexible numerology feature, 5G-NR can enable extremely high data rates everywhere, extremely low latency, ultra-high reliability and availability, extremely low computing cost and energy usage, and energy-efficient networks.

\item {\bf Flexible Spectrum Formation:} 5G will support two frequency spectrum ranges, namely Frequency range 1 (FR1) (below 6 GHz) and Frequency range 2 (FR2) (above 24 GHz). 5G is expected to operate in unlicensed spectrum (such as 2.4 GHz, 5 GHz and 66 GHz) while co-existing with technologies such as WiFi. Moreover, discussions are ongoing to allocate a dedicated spectrum for 5G based vertical industries \cite{GSMA_2020}. In comparison to LTE, where maximum channel bandwidth is 20 MHz,  the maximum channel bandwidth per 5G Carrier is 400 MHz. For 5G NR, the frame length is set at 10 milliseconds, the length of a subframe is 1 milliseconds, the number of subcarriers per resource block (RB) is 12, and each slot contains 14 orthogonal frequency-division multiplexing (OFDM) symbols (12 symbols for extended cyclic-prefix mode), all of which are identical to LTE. In comparison to LTE numerology, which uses a 15-kHz subcarrier spacing, the 5G frame structure allows for subcarrier spacings of 15, 30, 60, 120, or 240 kHz. The features above (i.e., flexible spectrum formation) are particularly important to support a wide range of data services with distinct QoS requirements in terms of reliability, latency, and data rate. 

\item {\bf Channel Coding:} In contrast to LTE, which uses convolutional codes and Turbo codes, 5G NR uses two capacity-approaching channel codes: low-density parity-check (LDPC) codes and polar codes, the former for error correction in user data  and the latter for control channels. Since both codes exhibit ultra-low decoding latency, they are particularly suitable in URLLC applications in smart manufacturing. 

\item {\bf Millimeter-Wave Spectrum (mmWave):} Currently, most of the wireless technologies use spectrum in the 300 MHz to 6 GHz band. 5G wireless networks will provide ultra-high-speed communication links by exploiting the unused high frequency mmWave band, ranging from 30 - 300 GHz. The cell size for 5G systems at mmWave frequencies is expected to be around 200 m \cite{Rappaport2013,Hur2016}. Through mmWave communications, 5G is expected to support augmented reality (AR) /virtual reality (VR) like advanced use cases of industrial environments, where very high data throughput is required.

\item {\bf Massive MIMO \& Smart Beamforming:} Using mmWave frequencies, due to the small wavelength, hundreds of antenna components can be arranged in an array on a small physical surface. For the mmWave bands, typical antenna numbers under consideration for the base station range from 256 to 1024. The arrangement of large number of antennas in 5G allows using an advanced access technology called Spatial Division Multiple Access (SDMA), which increases the capacity of the system to support a large number of sensor nodes simultaneously \cite{Gupta2015,Agiwal2016,Hong2017} in a smart factory.

\item {\bf Signal Processing for 5G:} With channel state information made available at the transmitter, analog based linear pre-coding (e.g., Matched Filtering) can be used for multi-user (MU) MIMO downlinks. However, this precoding technique may suffer from high complexity for massive MIMO scenarios. In recent years, hybrid precoding (combination of analog and digital precoding) emerged as a potential solution for 5G MU-MIMO downlinks, especially with mmWave enabled massive MIMO \cite{Shafi2017,Andrews2014,Gao2015,Alkhateeb2014}. With massive MIMO, linear processing such as MMSE/ZF \cite{Hoydis2013,Bjornson2016} can be used as interference mitigating receivers in the uplink of 5G. However, in recent years, reduced complexity non-linear methods \cite{Roh2014,Hossain2014,Nam2014} have been developed to work with mmWave and massive MIMO enabled 5G. When used in a dense network environment, the advanced signal processing technique in 5G is expected to  boost communication quality.

\item \textbf{Software Defined Networking (SDN):} Software-defined networking facilitates network management through a softwarization approach, in which the data plane is separated from the control plane to improve network performance and monitoring. A separation between control plane and user plane functions in 5G core network leads to the following Functional Entities (FEs): Mobility Management Control Function (MMCF), Session Management, Control Function (SMCF), Policy Function (PF), Subscriber Database Function (SDBF), Authentication Function (AuF), Application Functions (AF) and User Plane Function (UPF) \cite{Shafi2017}. A new interface, so called NGx has recently been defined by 3GPP for interaction between these key functional entities. 

\item \textbf{Network Function Virtualization (NFV):} Virtualization of network functions arose in the data centre community, with the aim of sharing common physical resources such as computation, storage, and networking by creating virtual machines (VMs). By implementing the same features in software, NFV decouples physical network functions (e.g., firewalls, routers, load-balancers, etc.) from dedicated hardware \cite{Mekikis2020}. It is  important to note that NFV and SDN are complementary technologies that meet the needs of demanding applications while optimizing physical network infrastructure utilization. 

\item \textbf{Network Slicing (NS):} Network slicing is a relatively new paradigm that allows for the creation of several logical networks customized to various types of data services and business operators. This feature in 5G will enable individual design, deployment, customization,
and optimization of different network slices on a common
infrastructure. Although the idea of NS was originally proposed for partitioning Core Networks using techniques such as NFV and SDN, it has since been expanded to provide effective end-to-end data services by slicing radio resources in Radio Access Networks (RANs) \cite{Li2016,Ordonez-Lucena2017}. Recently, a 5G-based network slicing framework has been applied to smart factory scenarios to interconnect different industrial sites and to provide the required end-to-end connectivity and processing features \cite{Taleb2019}.

\end{itemize}

The above features of 5G wireless network could accommodate different use cases running in the same smart manufacturing environment with the aid of the flexible sub-carrier formation and network slicing. For example, a Human-Machine Interaction application with high and constant data rate could efficiently share the radio resources with a Collaborative Machines applications with synchronized and low latency cyclic operations. Each radio resource slice allocated for a use case may use a different numerology (sub-carrier spacing and time-slot duration) to meet its end-to-end connectivity requirements and they can coexist in the same radio interface.
The use cases that require massive connectivity (i.e., a large number of deployed devices such as sensors to perform Predictive Maintenance or Remote Operation activities), are significantly benefited by the use of the wide mmWave spectrum. Although wireless sensor networks usually generates narrow-band and low rate traffic, the lower frequency bands may not support simultaneous connectivity of thousands of devices. 
Moreover, the use cases that demands very high throughput such as Human-Machine Interaction applications are also benefited by the mmWave wide band and by the high spectral efficiency of Massive MIMO \& Smart Beamforming techniques. 5G prototype networks have already been deployed for industrial applications. For example,  Qualcomm has designed the network architecture and communication protocol to allow 5G to operate as part of existing Time Sensitive Networks (TSN), which are deployed in an industrial space \cite{Khoshnevisan2019}. Qualcomm has developed a prototype to validate the proposed design and demonstrated its  performance by tracking and controlling high speed rotating discs which are used in industry. It is shown that 5G communication is able to achieve similar performance as  ethernet cable communication in terms of data throughput.




\subsection{WIFI 6 (802.11ax)}
\begin{figure}[b]
    \centering
    \includegraphics[width=\linewidth]{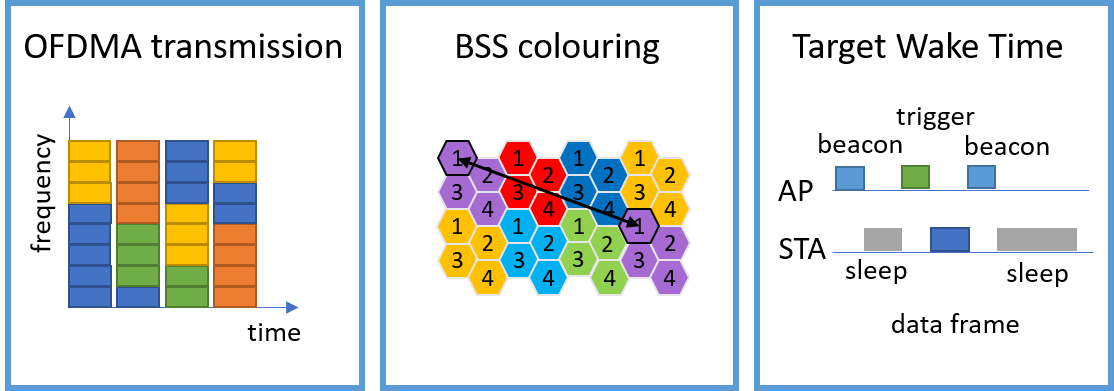}
    \caption{New Wi-Fi 6 features for capacity enhancement.}
    \label{fig:wifi6features}
\end{figure}
The WLAN standard IEEE 802.11ax, marketed as WI-Fi 6 by Wi-Fi Alliance \cite{WiFi6},
enters the market of industrial wireless automation and digitalisation, aiming to increase the number of WLAN clients, the spectrum efficiency, and the runtime of battery-powered devices.
This can be mainly achieved with finer radio resource allocation, spatial reuse and a new wake-time mechanism, given the following main features:

\begin{itemize}
    \item \textbf{Orthogonal Frequency Division Multiple Access (OFDMA)}: Wi-Fi 6 uses the same spectrum allocation on 2.4 GHz and 5 GHz as previous versions with sets of 20 MHz wide channels that can be aggregated in blocks up to 160 MHz wide.
    However, within those 20 MHz channels, Wi-Fi 6 subdivides the frequency space into 256 sub-channels - four times more than the 64 sub-channels previously used.
    This improves the resolution with which a link can cope with interference, frequency-selective fading and resource allocation.
    Those sub-channels are organised in an Orthogonal Frequency Division Multiple Access (OFDMA) transmission frame, where a communication channel is divided into up to nine sub-channels, a so-called  resource unit (RU). These RUs can be allocated to different clients communicating simultaneously.
   In the previous version, they were organised in an Orthogonal Frequency Division Multiplex (OFDM), where only one client can communicate at a given time since the full channel is used during the data transmission.
    This means Wi-Fi 6 can achieve much lower latency and fairer distribution of bandwidth between clients as it can serve multiple users in less time.
    Also, it can provide more efficient transmission of small data packets as the transmission can be allocated into a small portion of the channel and does not waste the whole bandwidth to transmit every time.
    

\item \textbf{Spatial Reuse}:
    A major cause of slow data rate in dense access point environments in previous Wi-Fi versions is the mutual interference between access points that share the same channel (co-channel interference) or whose channel groupings overlap (overlapping basic service set).
    Using the Carrier Sense with Multiple Access Collision Avoidance (CSMA/CA) mechanism, a radio wanting to transmit first listens on its channel, and if it detects an ongoing transmission, it backs off before trying again.
    In Wi-Fi 6, the interference between relatively close access points (using the same channel but different BSS) is mitigated by assigning different ``colours" for each BSS.
    Before transmitting, an access point or a client detects a signal on its channel and checks the colour code. If the colour is different and the signal strength is low enough (indicating a low chance of interference), the transmission is done without the need for backing off.
    Hence, the same channel can be reused by different access points separated by a distance that gives enough signal attenuation.
    Therefore, spatial reuse enables a more efficient use of the spectrum if the access point deployment is adequately planned.
    

    \item \textbf{Target Wake Time (TWT)}:
    The TWT mechanism introduced by Wi-Fi 6 aims for improving the battery life of IoT sensors and other devices. This is possible by allowing the devices to ``sleep" between sending and receiving packets and waking up at an agreed time, which is negotiated between the access point and the client.
    During the sleep time, the device enters a low-power mode and consequently saves battery.
    The wake time can be scheduled by the access point to create efficient spectral usage patterns and to maximise the number of clients it can handle over time.

    \item \textbf{Other Improvements}: Wi-Fi 6 standard also introduces other features that improve the general WLAN performance, such as uplink Multi-User Multiple Input Multiple Output (MU-MIMO) and higher modulation scheme 1024-QAM, increasing the nominal data rate.
    
\end{itemize}

Wi-Fi 6 can enable use cases such as massive sensor network deployments and Human-Machine Interaction with Augmented/Virtual Reality (or Mixed Reality) applications with these enhancements.
As described in the previous sections, factories can be equipped with many sensor devices that help predictive maintenance, manufacturing operations, theft prevention, and logistics monitoring.
All these devices can be potentially connected to a single Wi-Fi 6 network supported by the increased client capacity and energy efficiency.
Mixed reality is commonly employed in information visualization for machinery design, and maintenance \cite{juraschek2018mixed} and generates significant amount of data from high-definition graphics and meta-data.
For instance, a ``walk-by" machine monitoring using a mixed reality application in a tablet can get an instant reading of real-time information of the machine \cite{wifi6trialMettis}.
However, such real-time applications might have performance degraded when client mobility increases.
In a dense and mobile industrial environment, clients might need to roam (handover) between different access points as they move.
The Wi-Fi 6 standard does not introduce improvements to the roaming process.
Currently, the client's roaming time can be between a few milliseconds and several seconds \cite{siemensWifi6whitepaper}, which may be unacceptable for critical applications and strict latency requirements.

\subsection{WIFI 7 (802.11be)} 
Wi-Fi 7 (IEEE 802.11be which is also named Extremely High Throughput (EHT) Wi-Fi) is a new amendment to the Wi-Fi standard that promises to scale up the nominal throughput to as high as 40 Gbps.
This is achieved mainly by enhancing the PHY layer (EHT PHY): doubling the bandwidth to 320 MHz, increasing the modulation to 4096-QAM, and increasing the number of spatial streams to 16$\times$16 MU-MIMO.
In addition to this primary goal, Wi-Fi 7 also aims to achieve higher spectrum efficiency, better interference mitigation, and real-time application (RTA) support.
However, the EHT PHY alone cannot provide the gains in throughput and latency to achieve all goals.
This is why other innovations are discussed for Wi-Fi 7: modified EDCA and OFDMA, multi-link operation for channel diversity, minimisation of channel sounding overhead, advanced PHY approaches (HARQ, NOMA and full-duplex), and multiple access point cooperation \cite{khorov2020current,deng2020ieee}.

        To support RTA, modifications are being proposed to the enhanced distributed channel access (EDCA).
        Many MAC solutions that are being discussed were initially proposed for wired Time-Sensitive Networks (TSN), e.g., interrupting a long delay-tolerant packet transmission to prioritise an urgent packet and scheduling transmissions \cite{tsnTaskGroup}.
        Other proposed solutions are introducing a new access category queue for RTA traffic \cite{adhikari2020proposals}, speeding up the backoff period of the RTA traffic \cite{zuo2019considerations}.
        Modifications in the OFDMA frame are also proposed as it allows the access point to manage downlink and uplink transmission centrally, and thus it can be a powerful tool to support RTA.
        Optimal allocation of resource units (RU) for an RTA client is needed, considering the knowledge of traffic parameters and packets' remaining lifetime, for example.
        Having such knowledge might require a cross-layer cooperation \cite{lai2018cross}, which might trigger substantial paradigm changes in the 802.11 standards.
        
        Wi-Fi 7 enables clients to seamless roam between access points in dense deployments by introducing a multi-AP cooperation feature.
        This technology allows for fast association and re-connection with an access point as users move around \cite{tanaka2019discussion}.
        It has been discussed that nearby access points will require cooperation with each other by coordinating the channel access and transmission schedule or by jointly transmitting the same data \cite{vermani2018terminology}.

Because of the mobility support, some use cases will be more benefited by Wi-Fi 7 than Wi-Fi 6, particularly in cases where there are dense access point deployment and robotic tools and autonomous intelligent vehicles are moving.
Once moving robots can seamless roam, productivity can be boosted by allowing robots to perform complex and dangerous tasks. These tasks may demand high data rate, ultra-low latency, and mobility to support the manufacturing process's synchronisation and management by offloading their potentially heavy computational workload to edge servers.
    
\subsection{6G} 
The Sixth Generation (6G) of cellular networks envisions to fully integrate advanced features that are only partially required in 5G technology, including artificial intelligence, autonomous vehicles, mixed reality and haptic interface technologies requiring low latency communication, which are also crucial features for Industry 4.0 and beyond \cite{IRS_2022}. To achieve this integration, the 6G wireless systems will indeed require more enhancements on throughput (up to 1 Tbps), network capacity ($\times$1000 5G capacity), energy efficiency, backhaul and access network congestion and data security.
The following key technologies enable these requirements \cite{chowdhury20206g}:
\begin{itemize}
    \item \textit{Artificial Intelligence}: The most important technology introduced for 6G systems is the integration of AI into the communciaion protocols of the future.
    AI will play a vital role in automatising complex and time-consuming network tasks, such as handover, network selection and channel access.
    Hence, it will increase efficiency and reducing delays in communication.
    \item \textit{Terahertz Spectrum}: The allocation of THz bands will increase the bandwidth available to support very high data rates. The high frequency will lead to high path loss, and very narrow beam antennas will likely be required \cite{xia2019expedited}.
    \item \textit{Optical Wireless Communication}: In addition to the variety of available RF-based communications, OWC technologies such as visible light communication (VLC), optical camera communication, and free-space optical (FSO) are considered for possible device-to-device, and fronthaul/backhaul networks \cite{chowdhury2019role}.
    \item \textit{Blockchain}: Managing a massive amount of data can be a bottleneck in the network if carried out on a centralised node. Blockchains are a form of distributed data management providing interoperability, security, privacy, reliability, and scalability \cite{dai2019blockchain}.
    Hence, blockchain technology approaches will be useful in the management of massive and secure Industrial IoT networks.
    \item \textit{Wireless Information and Energy Transfer}: WIET allows the system to transfer wireless power during communication using the same electromagnetic wave.
    Hence, battery-powered sensors can be charged while communicating the data, thus lengthening the devices' lifetime.
    Also, network devices without batteries could be supported.
    \item Other key enabling technologies include: integrated airborne networks (UAV, satellite) \cite{mozaffari2018beyond,giordani2020satellite}, quantum communications \cite{nawaz2019quantum},
    integrated sensing and communication \cite{kobayashi2018joint}, enhanced MIMO techniques \cite{chowdhury20206g} .
\end{itemize}

Although the transformation to Industry 4.0 will be enabled by secure, low latency wireless communications and automation through the use of 5g and the other established protocols described, future industrial use cases with much more extreme requirements will benefit from the advancements associated with 6G \cite{viswanathan2020communications}:
\begin{itemize}
    \item The use of augmented/virtual reality in industrial tasks may benefit from higher resolution and multi-sensory designs.
    \item Massive deployments of mobile robot swarms and drones performing a vast range of tasks may benefit from increased capacity and link reliability and distributed computing.
    \item Dynamic digital twins may benefit from increased accuracy for synchronous updates from the physical world and higher resolution of real-time mapping and rendering.
\end{itemize}

\section{Challenges and Future Directions} \label{sec:Future_Works}
\subsection{Wireless Power Harvesting}
As the sensor nodes supported by most of the wireless technologies are battery powered in nature, any form of harvesting energy from the ambient environment could help achieving an energy optimal operation and reduce maintenance costs associated with battery operated devices. Wireless energy harvesting is a promising and widely adopted solution to provide power to wireless senor nodes installed in a harsh industrial environment (i.e., inside the machinery on a production line) where the total-cost-of-ownership can be significantly increased by the need to replace batteries.

Despite the recent advances in the area of wireless power harvesting, there are still a number of issues that need to be addressed. Over the past number of years, research activities have focussed on the hardware implementation and transmission management mechanisms with respect to physical layer in wireless power harvesting networks \cite{shaikh2016energy}. However, there are several issues related to integrating RF-power harvesting into the existing communication protocols as regards how the data and energy transmission mechanisms can dynamically be adjusted to share the same spectrum. This subsection summarises those challenges as well as potential solutions in an effort to provide guidelines and research directions to work on for the research community.

One challenge being addressed by the research community is in the field of energy harvesting and utilization efficiency in wireless powered harvesting networks. In such energy harvesting techniques, RF energy is impacted on by spreading loss in the free space and the resultant energy received by the end device in the electromagnetic far-field is very limited \cite{pinuela2013ambient}. Hence, both the hardware (such as the transmitters, the harvesting circuit as well as antennas) and the software (e.g., the power management policy and the
transmission strategy) should be designed to maximize the energy harvesting transfer, energy conditioning and energy utilization efficiency of the overall systems including sensing and communications

Establishing the optimum tradeoff in relation to the interference between the wireless energy transfer mechanism and the wireless communications is another critical challenge in wireless powered networks. In the case of both dedicated and ambient power harvesting sources, the interaction between energy transfer and communications is significant as the energy transmission can interfere with the information decoding while data transmission may cause interruption in energy reception while operating in the same bands. Hence, the opimisation of the power transmission and data throughput is a challenging design issue \cite{zhao2017exploiting}. In this case, a softwarized on/off mechanism at the MAC protocol level is inevitable in order to separate the energy and data slots to ensure no communication is taking place when the end devices are being charged. As such, wireless power harvesting requires a scheduling policy to manage interference \cite{hadzi2017opportunistic}. Spectrum scheduling could be a useful tool to avoid or mitigate the interference by employing some management techniques such as interference cancellation and alignment to optimise energy transfer and communications.

Network coding schemes could be another factor heavily influencing the energy efficiency of such wireless sensing systems. These allow the senders to transmit the information and energy simultaneously and improve the RF energy harvesting efficiencies, particularly in the case of large-scale networked systems. The senders and relays can effectively use available time slots for power harvesting when they are idle from a communications perspective. The study \cite{jungnickel2014role} suggests that network lifetime gain could be improved up to 70\% when employing RF power harvesting. A diverse range of network model and coding schemes need to be explored in order to improve the network lifetime. However, it is still an open issue to investigate whether the RF power harvesting improves the upper bound of energy gains and if so, to what extent.

Size constraints in the sensor design are also a challenge which needs to be addressed in many wireless sensing applications \cite{rezaei2020large}. The antenna, matching network, and the rectifier are the integral components of an RF power harvester and the antenna size has a decisive role in determining the harvesting density so it becomes even more pertinent to reduce the size of the devices embedded to the designed sensors while maintiaing harvesting efficiency associated with larger antenna.

Energy trading strategies could be another dimension in wireless power harvesting networks \cite{chen2017real}. For instance, wireless charging service providers could be seen as small RF power suppliers contributing towards (partially) meeting the energy demands of nodes throughout the network. Seeking the best trade-off between the amount of RF power harvesting and the price to optimize the revenue and costs should always be considered.

Finally, the health impacts of wireless energy harvesting are an area which causes some concern in certain research circles. It has widely been acknowledged that exposure to electromagnetic waves at certain frequencies causes heating some materials with finite conductivity \cite{code2015limits}. Several studies \cite{wood2006dangerous,ahlbom2008possible,hardell2008biological,breckenkamp2009feasibility,habash2009recent} have been conducted on the effects of electromagnetic radiations from mobile devices and cellular infrastructures and a majority of these studies mark the radio waves as non-harmful for human health. Only a few studies \cite{breckenkamp2009feasibility,habash2009recent} report some effects on genes when exceeding the upper bound of internationally recognised standards on secure power levels. little research has been carried out to investigate the impact of dedicated wireless RF chargers on health which can sometimes emit much higher powers than those used for communications. As such, there is a need to gauge the appropriate safety standards while deploying these RF chargers.

\subsection{Millimetre-Wave and Terahertz based Industrial Communication}

The millimetre-wave and Terahertz (or sub-millimetre-wave) spectrum have large bandwidth available to provide multi-Gbps throughput for the upcoming industrial applications.
Besides communication, these frequencies can also be used in industry for accurate positioning of artefacts and material sensing.
Object positioning is benefited as more bandwidth is available compared to lower frequencies, leading to higher accuracy and lower latency  \cite{lu2018opportunities}.
These frequencies can also be used to scan materials for product quality inspection without opening the package or disturbing the content, measuring moisture levels, layer thickness, and surface uniformity \cite{scott2010terahertz}.

The use of mmWave and THz frequencies in industrial environments opens new challenges that were not considered using lower frequencies since path loss, and other channel impairments become more significant than in lower frequencies.
As frequency increases, the wavelength decreases, and consequently, more objects act as scatterers.
Small changes in the path length will cause small-scale fading, and obstructions of the line-of-sight signal will cause large-scale fading effects (blockage) \cite{cheffena2016industrial}.
In an industrial environment, building and furniture materials are commonly made of concrete and metal, reflecting the radio waves and increasing the scattering of the signal.
Furthermore, the movement of workers, machinery and other objects may cause a dynamic variation of the channel conditions \cite{cheffena2012industrial}.
Hence, the type of industrial environment has a significant impact on channel propagation and link performance. 
For example, in the ``light industry", where the production is end-consumer oriented, the products and equipment have a small impact on the propagation.
While in ``heavy industry", where the machines are larger and heavier, the setup complicates the propagation \cite{solomitckii2018characterization}.
Therefore, a mmWave/THz network deployment has to be carefully designed to mitigate the channel impairments according to the type of industrial environment.

To overcome the issue of path loss, narrow beam antennas are used.
However, considering a massive user deployment, the beamwidth design can lead to a fundamental trade-off between the area covered (and the number of users) and latency \cite{mazgula2020ultra}.
On the one hand, with wider beamwidth, the antenna gain is lower and, consequently, the SNR is reduced, leading to some users without enough SNR for successful transmission. However, with a wider beam width, more users can be illuminated by the beam and be served simultaneously, decreasing resource allocation's latency.
On the other hand, with narrower beamwidth, the SNR increases, leading to an increased data rate despite serving fewer users. Besides, reflected/non-line-of-sight paths become weaker, resulting in more gaps in coverage.

When using the same spectrum for both communication and THz sensing in the same environment, there are inevitable compromises that must be made.
For example, in a mmWave system, the position can be estimated based on a pilot signal that is sent together with the data signal \cite{han2020millimeter}. The accuracy of the position measured increases with the number of pilot signals sent, which takes resources from the data signal, decreasing the data rate.
Furthermore, it is not desirable that the communication directly interferes with the data gathering and vice-versa.
Therefore, mechanisms shall be designed to isolate (allocation in time or space) the process from each other.

\subsection{Guaranteed Low Latency \& High Reliability}

Guaranteed low latency and very high reliability are two requirements of Industry 4.0 communications \cite{varghese2014wireless}. Both requirements play an important role in the pace of productivity, management of production facilities, and as a consequence, in the maximization of profit. 

A number of MAC protocols have been developed to achieve high reliability and low latency using existing radio technologies. IPv6 over Low-Power Wireless Personal Area Networks (6LoWPAN) standards such as the WirelessHART and the 6TiSCH make use of time-slotted MAC with channel hopping capabilities over the well-known IEEE802.15.4 radio interface. Both protocols, achieve very high reliability mainly due to the channel hopping mechanism which allows packet re-transmissions over non-repeated channels. Thus, transmissions that previously failed due to interference at a specific channel, have more chances to be delivered by using another pseudo-random channel in the next attempt. The time-slotted attributes of the MAC assures deterministic latency between 10ms to 1s depending on the length of the slotframe \cite{ietf-6tisch}. The biggest disadvantage of the IEEE802.15.4-based protocols is that they operate at the 2.4GHz ISM frequencies which may be overcrowded by other dominant radio technologies, mainly the IEEE 802.11 ones. Due to their higher transmit power, IEEE 802.11 signals usually dominate the IEEE 802.15.4 and lead to a burst of packet losses \cite{sikora2005coexistence} and a degradation of the reliability. Nevertheless, recent studies propose to blacklist badly-performed channels in order to increase the reliability levels \cite{kotsiou2019blacklisting}. How to efficiently achieve blacklisting throughout a multi-hop network with the minimum possible overhead is an open problem.

Even though 6LoWPAN present good performance and have been advanced quite well in the standardization process, they cannot easily support long range applications due to the short range nature of their radio. Multi-hop topologies add additional delay, cost, and network overhead because of the activity of the routing layer protocol. Researchers have studied the potential of LPWAN radio technologies for industrial applications. For example, Zorbas \textit{et al.} propose a LoRa-based time-slotted protocol with deterministic delay and high reliability \cite{ZORBAS2020}. The authors target applications that require up to a few seconds of delay in between successive transmissions such as predictive maintenance, building health monitoring, and asset tracking applications. Other researchers explore the possibility of using a time-slotted version of IEEE 802.15.4-Wi-SUN radios \cite{munoz2018overview}. However, sub-GHz transmissions suffer from very low radio duty cycles given the existing regulations across the world, which limits the minimum latency that the solutions can achieve.

However, there are many applications that require a very fine synchronisation between devices that cannot be achieved with existing radio technologies. For example, robot motion control requires latency of less than 1ms and support of hundreds of devices in a 10 km$^2$ industrial area \cite{brown2018ultra}. Thus, researchers have recently focused on the URLLC capabilities of 5G or other mmWave technologies to address diverse, high-performance use cases linked to industrial automation as it was mentioned in Section \ref{sec:Emerging_Tech}.


\begin{figure}[h]
\centering{\includegraphics[width=85mm]{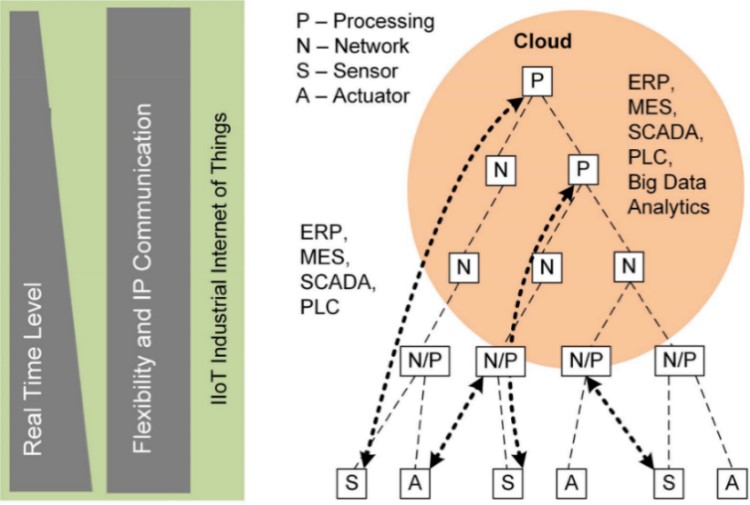}}
\caption{Industrial IoT Paradigm \cite{Bruckner2019}.}\label{fig:IIoT_Paradigm}
\end{figure}

\subsection{Deployment and Interoperability}
Today, communication solutions in factories are heterogeneous and often not interoperable. With the advent of Industry 4.0, the requirements are defined for industrial production systems such as mass customization and efficient automated production lines for small lot sizes. These requirements are mainly data-driven cloud-based services that require a more flexible, automated architecture \cite{Wollschlaeger2017}. Industry 4.0 builds on the IoT paradigm that suggests a flat cloud of interconnected devices rather than a complex hierarchical model \cite{Bruckner2019}. Applying the Internet of Things (IoT) paradigm leads to a new automation architecture, called Industrial IoT, as shown in Fig.~\ref{fig:IIoT_Paradigm}. The new architecture calls for uniform communication across all functional levels based on IP and thus fewer gateways are needed (less configuration effort). In this case, the typical automation communication requirements such as high reliability, low latency, and time synchronization, high amount of data and devices as well as seamless reconfiguration, sensor-to-cloud communication \cite{Kobzan2018}, and convergence are present and addressed on all functional layers \cite{Heymann2018}. To address the new requirements of IIoT, adequate communication systems are demanded wherein IT and OT converge. In the same context, the IIoT vision requires the field devices to support IP, which is not possible with legacy Fieldbus systems and still partly difficult with the various existing real time Ethernet solutions \cite{Bruckner2019}. Therefore, a further evolution of industrial communication systems is required. 

Time Sensitive Networking (TSN)  \cite{Finn2018} is a recent standardization activity, which has the potential to harmonize today’s industrial communication systems, which realizes the IIoT requirements \cite{Pop2018}. Most of the TSN standard elements have been primarily proposed for wired Ethernet networks. However, recent efforts try to extend TSN to wireless technologies \cite{Bush2018}, \cite{Cavalcanti2019}, including WiFi \cite{Stanton2018} and 5G \cite{Mahmood2019}, which will provide a vendor-independent solution that can achieve full interoperability.  The integration of TSN and wireless network has the potential to achieve the first unified network stack for the industrial communication, which promises to accomplish all requirements of Industry 4.0 and the IIoT architecture \cite{Bruckner2019}.

\section{Conclusions} \label{sec:conclusion}

A wide range of wireless technologies for machine-to-machine (M2M) communication for factory or process automation applications have been developed, standardized or proposed. These include low power short range wireless communications and network technologies such as WiFi5/6 and WIA-FA, which are based on the IEEE802.11 standard, or WiHART, ISA100.11a, WIA-PA, based on the IEEE802.15.4 standard to IO-Link, which is based on the IEEE802.15.1 standard. Some of these technologies cover only the lower layers of the protocol stack, such as the PHY and MAC layer (WiFi, WIA-PA/FA) and typically use Internet protocol layers or automation protocols such as OPC-UA for higher layer functions. Others are full network protocol stacks such as WiHART, ISA100.11a or IO-Link. Some of these short-range technologies such as WiHART or ISA100.11a are already deployed in many manufacturing environments and so are WISA and WISAN-FA, predecessors of IO-Link. Short range wireless technologies typically target indoor factory environments or confined outdoor areas.
In contrast, M2M technologies that are based on low power wide area technologies such as LoRa/LoRaWAN, NB-IoT or LTE-M will likely target outdoor plant areas due to their wide area coverage. These technologies also implement only lower protocol layers and will likely either use proprietary higher layer protocols or will leverage open protocols such as Internet protocols or OPC-UA to integrate them into automation and control applications.

Recently, 5G wireless technologies are being actively standardized and are seeing first deployments in the consumer space. 5G is a full network standard and is seen as the future wireless network platform for a very wide range of vertical applications, including factory automation. In particular, mMTC and URLLC services specified as part of the 5G standard aim at smart manufacturing use cases. The former aims at dense wireless sensor and actuator deployments, whereas the latter aims at supporting robotics and autonomous vehicles in factory environments. While 5G standardization is progressing, there are currently no products available on the market that support mMTC or URLLC applications. Some wireless manufacturers have started to evaluate 5G technology for automation applications in laboratory environments, but there is no experience with large-scale deployment available. While 5G wireless offers very flexible wireless access, it is unclear how energy efficient it is for dense low power sensing and actuation applications in manufacturing, e.g. thousands of sensors on one production line.

While there is much hype about 5G in a broad range of areas including manufacturing and logistics, some earlier mentioned wireless technologies are finding and will continue to find application. This will require future factory environments to accommodate heterogeneous wireless technologies comprising  short-range and wide-area wireless as well as future 5G technologies. This heterogeneous environment will also need to cater for the planned integration of Information Technology (IT) and Operating Technology (OT) in manufacturing, which has so far been quite separate but will merge as part of the Industry 4.0 vision.
Future wireless factory scenarios will likely see a range of wireless technologies in use, all connected by a factory backbone communication network that caters for both IT and OT applications. Such a backbone could be based on a slice of the 5G core network or could be based on a combination of private 5G network and other emerging network technologies such as TSN, a new Ethernet based real-time network technology targeted among others at factory automation environments.

From this survey, it is clear that much research is still required to explore the use of wireless technologies in smart manufacturing. So far wireless technologies have seen limited deployment in factory automation application and many questions as to their reliability, dependability, performance and security remain open. The variety of available wireless technologies all aim at a specific subset of manufacturing use cases. However, it is unclear what works best for which case. 5G has attracted much hype and is seen by many analysts as the future for wireless factories. However, there is currently no experience of the performance of 5G in factory automation due to the lack of products, deployments, and large-scale studies. In order to advance the application of wireless technologies in factory or process automation, more research is required that explores issues such as powering devise through energy harvesting, performance of wireless technologies in mmWave and Terahertz bands, ultra-reliable low latency issues, and the design and deployment of these technologies both in simulated environments in a first step and then in real deployments in prototype or real manufacturing or factor environments.

\section{Acknowledgements}
This publication has emanated from research conducted in part with the financial support of Science Foundation Ireland under Grant number 16/RC/3918 (CONFIRM Centre for Smart Manufacturing) and the Johnson \& Johnson Advanced Technology Centre.

\bibliographystyle{IEEEtran}
\bibliography{references}

\end{document}